\documentclass[11pt,twocolumn,A4]{article}
\usepackage{psfig}
\usepackage{amssymb}
\usepackage{a4}
\usepackage{fleqn}
\newcommand{\beq}{\begin{eqnarray}}
\newcommand{\eeq}{\end{eqnarray}}
\newcommand{\punkt}{\, .}
\newcommand{\komma}{\, ,}
\newcommand{\quadbox}[1]{\qquad\mbox{#1}\qquad}
\newcommand{\romg}{\mbox{\rm g}}
\newcommand{\romf}{\mbox{\rm f}}
\newcommand{\romerf}{\mbox{\rm erf}}

\newcommand{\mmm}[1]{\mbox{$\langle M \rangle_{#1}$}}
\newcommand{\mean}[1]{\mbox{$\left \langle #1 \right \rangle$}}

\newcommand{\sigmas}{\sigma_s}
\newcommand{\sigmaf}{\sigma_f}
\newcommand{\sigmafhat}{\hat{\sigma}_f}
\newcommand{\ynorm}{\mbox{$\frac{\bar{H}_s}{\sigma_s}$}}
\newcommand{\norms}{\mbox{$\displaystyle \frac{2}{\sqrt{\pi} \sigmas
\left( 1+{\rm erf} \left( \ynorm \right)\right)}$}}

\newcommand{\Hex}{\mbox{$H_{ext}$}}
\newcommand{\hex}[1]{\mbox{$H_{ext}^{(#1)}$}}
\newcommand{\hiup}[1]{\mbox{$H_{i,\uparrow}^{(#1)}$}}
\newcommand{\hidown}[1]{\mbox{$H_{i,\downarrow}^{(#1)}$}}

\newcommand{\wdd}{\mbox{$w_{\downarrow\downarrow}$}}
\newcommand{\wud}{\mbox{$w_{\uparrow\downarrow}$}}
\newcommand{\wdu}{\mbox{$w_{\downarrow\uparrow}$}}
\newcommand{\wuu}{\mbox{$w_{\uparrow\uparrow}$}}
\newcommand{\thc}{$T_{Hc}$}
\newcommand{\Tc}{$T_C$}
\begin{document}
\onecolumn
%
\begin{titlepage}
\begin{center}
{\Large \bf Thermal Remagnetization in Polycrystalline Permanent
Magnets}\\[2ex]
by \\[2 ex]
{\large R. Schumann$^{\dagger}$ and L. Jahn $^{\dagger\dagger}$
}\\[2ex]
{$^{\dagger}$\it Institute for Theoretical Physics, TU Dresden,
D-01062 Dresden, Germany}\\
{$^{\dagger\dagger}$ \it Institute for Applied Physics, TU Dresden, D-01062 Dresden, Germany}
\begin{abstract}
The thermal remagnetization (TR), i.e. the reentrance of
magnetization upon heating in a steady-field demagnetized sample,
is a common feature to  the four types of polycrystalline
permanent magnets, mainly utilized for practical purposes, i.e.
barium ferrites, $SmCo_5$, $Sm_2Co_{17}$ and $NdFeB$ magnets. The
effect is small for pinning controlled and large for nucleation
controlled magnets. The effect is strongly dependent on the
demagnetization factor and may reach nearly 100\% in $SmCo_5$
samples measured in a closed circuit. The TR is very sensitive to
a small superimposed steady field. The maximum effect and the
position of the peak is dependent on the initial temperature. The
direction of the TR is correlated with the temperature coefficient
of the coercivity, resulting in a inverse TR in barium ferrite.
The susceptibility of the thermally remagnetized samples is
increased. Repeated cycles of steady-field demagnetization
followed by heating result in the same TR. The phenomenology of
TR and ITR is explained by means of a model taking into
account both the internal field fluctuations due to grain
interactions and the decay of single domain grains into
multi-domain state. By taking the measured temperature
dependencies of the coercivity and the saturation magnetization
the theory is able to reproduce the experiments very well,
allowing to determine the width of the field fluctuations, the
width of the switching field distribution and an internal
demagnetization factor as characteristics of the materials by
fitting.
\end{abstract}
\end{center}
\vspace{1cm}
{\bf KEYWORDS:} permanent magnets,  thermal remagnetization,
interaction
fields, switching field distribution \\[1ex ]
{\bf PACS:} 75.50Vv, 75.50Ww, 75.60Ej\\[1ex ]
{\bf CORRESPONDING AUTHOR:} R. Schumann \\[1ex ]
{\bf E-MAIL:}  {\it schumann@theory.phy.tu-dresden.de} \\[1ex]
\end{titlepage}
%
%
\twocolumn  
\section{Introduction}
The effect of Thermal Remagnetization (TR)  was first mentioned in
a paper of Lifshits, Lileev and Menushenkov \cite{Lifshits74}
devoted to stability investigations of the coercivity of
$SmCo_5$-magnets at elevated temperatures in 1974. They wrote: "We
also found a remagnetization upon heating  in absence of an
external field at samples, which were partially or completely
demagnetized by a steady counter field." \cite{Schumann03}. One
year later the same group \cite{Kavalerova75} published a first
systematic study devoted solely to this effect. As we will see in
the following, TR is not restricted to $SmCo_5$ magnets, instead
it is quite common to all sintered hard magnets. But, before
giving an outline of the history, we have to explain the principle
of the TR experiment.
\begin{figure}[!h]
\psfig{file=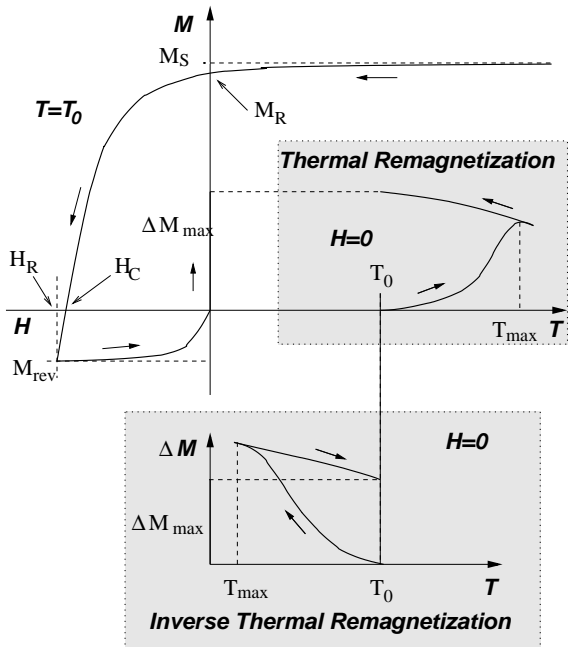,width=7.5cm}
\caption{The scheme of the standard TR-Experiment.
After saturation the sample is
demagnetized isothermally by a steady field at the initial
temperature $T_0$ (upper sketch) and afterwards either heated
(TR-experiment - left shaded picture) or cooled (ITR-experiment)
with the external magnetic field switched off, whereby the magnetization
is measured.}
\label{prinzip}
\end{figure}
In Fig. \ref{prinzip} the experimental procedure is sketched.
Firstly the sample has to be saturated by a very large magnetic
field. Next a steady field $H_R$ (remanence coercivity) opposite
to the saturation direction is applied. Switching off the external
field drives the sample in the dc-demagnetized state. Now the
temperature is changed and the related change of magnetization is
measured. If a reentrance of magnetism with increasing temperature
is observed the effect is called thermal remagnetization (TR), see
the right shaded picture. Otherwise, if a magnetization return is
observed upon cooling, as shown in the lower shaded figure, we
call it inverse thermal remagnetization (ITR). Already in their
first investigations Lileev et al. \cite{Lifshits74,Kavalerova75}
demonstrated for $SmCo_5$, that the TR is strongly dependent on the
demagnetization factor, especially it amounts up to nearly 100 \%
in a closed circuit. Furthermore it was shown, that at least
for $SmCo_5$, the effect is not necessary restricted to the
demagnetized state, but also occurs in a partially dc-demagnetized
state ($\pm$ 23 \% of the saturation magnetization). On the other
hand the effect is absent in ac-demagnetized samples.
Reconsidering the TR in $SmCo_5$ nearly ten
years later, Livingston and Martin \cite{Livingston84} suggested
as explanation of the TR the conversion of
the single domain state (SDS) of some grains into a multidomain state (MDS).
This assumption was supported
by the observation of multidomain states in a few big grains of
$SmCo_5$ at higher temperature and also by the discovery of
the TR in sintered $NdFeB$-magnets \cite{Livingston85}, where the
effect is larger in samples with lower number of multidomain
grains in the dc-demagnetized state. An alternate explanation was
given in Ref. \cite{Jahn85}. Based on investigations of the TR in
$SmCo_5$ and barium ferrite in 1985 the authors proposed the magnetic
interaction between adjacent grains together with a broad
distribution of the temperature dependent switching fields to be
responsible for the TR. This model, despite of its
oversimplifications, was able to explain the increase of
magnetization qualitatively and it showed the key role of the
temperature coefficient of the coercivity resulting in the
discovering of the ITR for sintered barium ferrites.
Furthermore it was demonstrated that the TR is connected with a
increasing susceptibility and that small external fields may
suppress the effect at all. Additional support the interaction model
got from a paper of Lileev and Steiner \cite{Lileev77}, proving
that the TR effect  depends strongly on the volume packing
density, a fact which was shown to be responsible for the
difference in ITR in sintered and chemical precipitated barium ferrite
samples \cite{Jahn85}. To decide, which of the both models
accounts well for the observed effects Lileev et al.
\cite{Lifshits76,Zaytzev88} counted carefully grains in single-domain
state and multi-domain state at different
temperatures. Their results proved that there is a delicate
interplay of both interaction and formation of multi-domain
grains, at least in the $SmCo_5$ samples. From the very beginning it
was assumed that TR is connected with the nucleation controlled
magnets. Otherwise, pinning controlled magnets should not show the
effect, and indeed, the $SmCo_5$-, $NdFeB$-, and ferrite-magnets show
remarkable TR and ITR resp.
\cite{Kavalerova75,Livingston84,Livingston85,Jahn85,Schumann87,Mueller88,Ivanov93},
 whereas the effect in $Sm_2Co_{17}$ is small
\cite{Scholl87}. To what extend this statement is true together
with the role of interaction in relation to the decay of SDS
grains into MDS grains, motivated much of the theoretical
and experimental work. Since the effect is most
conspicuous in $SmCo_5$, much effort was spent to this material.
Otherwise it became clear, that it is a rather common effect to
all modern poly-crystalline permanent magnets. Nevertheless, as will be
shown in the next section, the experiments demonstrate clearly the individuality
of every compound class.
Regarding the nucleation
controlled magnets we find that both the saturation magnetization
and the coercivity decrease monotonously for $SmCo_5$- and
NdFeB-magnets, but, while in $NdFeB$ magnets the coercivity and the
saturation magnetization vanish at nearly the same temperature, in
$SmCo_5$ the coercivity reaches zero at considerable lower
temperature than the saturation magnetization. In contrast
barium ferrite shows an increasing coercivity up to 550 K, i.e. the
temperature coefficient of the coercivity has the opposite sign. For
the $Sm_2Co_{17}$-magnets $M_S(T)$ and $H_C(T)$ behave as in $SmCo_5$, but
there we have a complete different coercivity mechanism, i.e.
volume pinning of domain walls. In $NdFeB$ magnets the magnetic active phase is
separated by one or more
other phases reducing the interaction between adjacent grains.
In Tab. \ref{table1} we show schematically
how the four groups of magnetic
materials are qualitatively distinguished by some features
being relevant for TR,
i.e. the number of phases,
the coercivity mechanism, and the temperature coefficient of
the coercivity.\\[1ex]
\begin{table}[t]
\small
\begin{tabular}{|l|p{1.2cm}|p{2cm}|c|} \hline
 Type          & \parbox[t]{1.5cm}{No. of\\ phases} & coercivity mechanism
                                      & \parbox[c]{0.8cm}{$\frac{dH_C}{dT}$}\\ \hline
SmCo$_5$       &      1        & nucleation           &  + \\
Sm$_2$Co$_{17}$& cellular      & pinning              &  + \\
NdFeB          &  $\ge$ 2      & nucleation           &  $+$\\
Ba-Ferrite     &      1        & nucleation           &  $-$ \\ \hline
\end{tabular}\\
\caption{Some of the main characteristics of the four groups of sintered permanent
         magnets.}
\label{table1}
\end{table}
In the third chapter, we will give  a brief review over the models used so far to
understand the different aspects of TR and present a theory which accounts for
both the TR in $SmCo_5$ and the ITR in barium ferrite in single-phase magnets.
The last chapter will be summarize what can be learned from TR measurements and
some remarks to open questions will be appropriate.
\section{Experiments}
\subsection{The standard TR experiment in the SmCo$_5$, Sm$_2$Co$_{17}$ and NdFeB}

\begin{figure}[htb]
\psfig{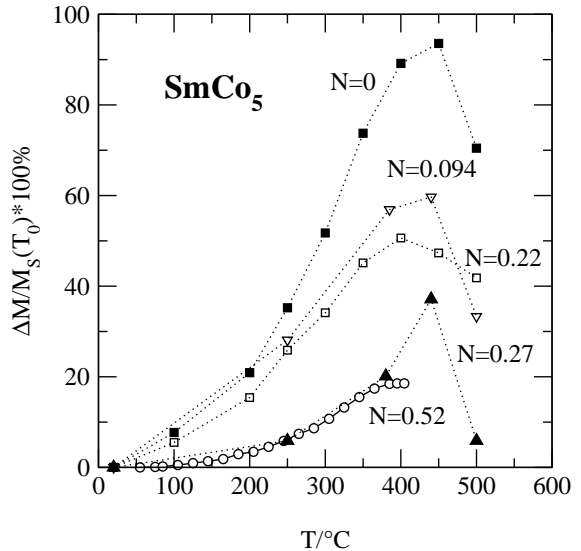}
\caption{The TR for various $SmCo_5$ samples taken from Ref.
\cite{Kavalerova75} (squares) and Ref. \cite{Livingston84} (triangles)
and a VACOMAX 200 sample (circle) \cite{Jahn02a}.
The data were transformed slightly from the original data to fit into
the same plot. Also the demagnetization factors of the cylindrical
samples were calculated from the geometry given by the authors. For details we
refer to the original papers \cite{Kavalerova75,Livingston84}.}
\label{SmCo5Kavalerova}
\end{figure}

Sintered $SmCo_5$-magnets for which TR experiments are reported
are single-phased and exhibit a poly-crystalline structure with the
typical grain sizes much bigger than the critical radius for
single domain behavior. They are mainly well textured.
The phase diagram \cite{Buschow68,Buschow73} of the Sm-Co system shows
that below about 750 $^{\circ}$C the magnetic
active phase $SmCo_5$ is meta-stable with respect to a
decomposition into $Sm_2Co_7$ and $Sm_2Co_{17}$ with reduced uniaxial anisotropy.
While at moderate
temperatures this decomposition process does not affect the
microstructure and the magnetic quantities, chemical instability becomes a problem
in high temperature
measurements, since a long lasting TR experiment may act as an
unintentional heat treatment of a sample.
The first measurements of the TR in $SmCo_5$ are collected in
Fig. \ref{SmCo5Kavalerova}.
The points shown as squares
are taken from Ref. \cite{Kavalerova75} and the triangles from
Ref. \cite{Livingston84} and the open circles are our own measurements
at a VACOMAX 200 specimen \cite{Jahn02a}.
The biggest TR (filled squares) was
measured in a closed magnetic circuit what corresponds to zero
demagnetization factor. The remaining curves represent open
circuit measurements at cylindrical samples of different aspect
ratio resulting in different demagnetization factors.
In Ref. \cite{Kavalerova75} the remanence was measured at room temperature
after heating to a certain temperature and cooling down to room
temperature, whereas in Ref. \cite{Livingston84} the magnetization
was registered simultaneously while increasing the temperature
stepwise. Every temperature step was followed by an annealing of
30 min. The remaining curves were measured while continuously
increasing the temperature. In order to get the slightly differing
experiments into one diagram we used the temperature dependence of
$M_S$ measured at our VACOMAX 200 sample together with the room
temperature remanence data given in Refs. \cite{Kavalerova75,Livingston84}
to scale the results.
Although one may object that the results of Fig. \ref{SmCo5Kavalerova}
stem from  $SmCo_5$ samples produced in different ways, it is evident
that a variation of the demagnetization factor
influences the TR much stronger than a variation of the production conditions,
at least if the result is a good $SmCo_5$ hard magnet.
\begin{figure}[htb]
\psfig{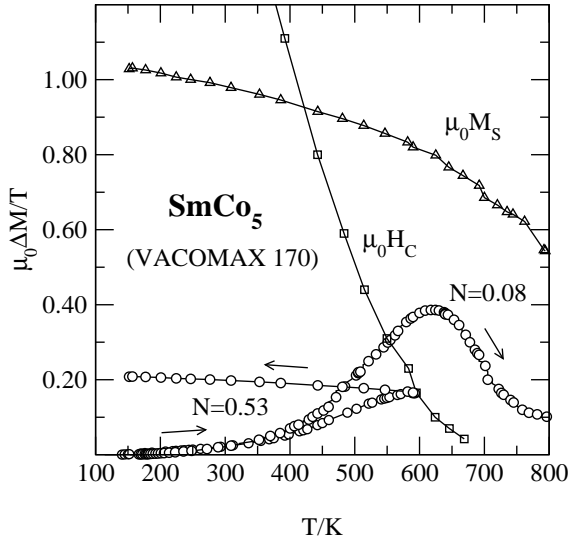}
\caption{The temperature dependence of the coercivity $H_C$, saturation
magnetization $M_S$ and the TR for two samples of $SmCo_5$ with demagnetization
factors N=0.08 and N=0.53 \cite{Jahn02a}.}
\label{SmCo5TR}
\end{figure}
Fig. \ref{SmCo5TR} shows both the TR for two samples with different
demagnetization factors cut from the same material (VACOMAX 170) \cite{Jahn02a},
and the temperature dependence of the saturation magnetization
and the coercivity of that material. It can be seen from the curve for the N=0.08 sample that
the TR survives in the region above the temperature \thc, where the
coercivity vanishes, which is about 700 K. As we will see in
the theory, this unexpected
behavior can be ascribed to the fact, that the Curie temperature
is more than 200 K higher than \thc.
Lileev et. al. measured the TR for $SmCo_5$ samples  with coercivities between (0.35 T/$\mu_0$ and
3 T/$\mu_0$. They found that the
inflection point, visible also in Fig. \ref{SmCo5TR}, vanishes for room temperature coercivities
\cite{Lileev92} below 1.6 T/$\mu_0$. For higher coercivities they found a decrease
of the initial slope and an increase of the maximum of the TR with increasing coercivity.
Measurements at samples, where the low-temperature
coercivity was reduced by a special heat treatment \cite{Rodewald} but otherwise prepared
in the same way as the samples utilized in Fig. \ref{SmCo5TR}, did not show significant changes
of the TR curves.\\

In contrast to $SmCo_5$, where the hysteresis mechanism is
nucleation controlled, the coercivity of $Sm_2Co_{17}$ magnets is
caused by volume pinning of domain walls. For that behave a
single phase solution of approximately 1:7 composition, i.e.
slightly Co deficient with respect to $Sm_2Co_{17}$, is homogenized at
approximately 1200 $^{\circ}$C and afterwards cooled down to 800
$^{\circ}$C, where it decays into the two adjacent stable phases by
precipitation of $SmCo_5$ and $SmCu_5$. The contemporary commercial magnets of the
2:17 type are composed typically Sm(Co,Fe,Cu,M)$_7$, where the
iron is used to increase the saturation magnetization and the
other metals mainly for metallurgical reasons to obtain the
precipitation microstructure. The latter is common to all
$Sm_2Co_{17}$-magnets. The entirely different mechanism of magnetic
hardening was the reason that in early years these compounds where
believed not to show TR.
\begin{figure}[hbt]
\psfig{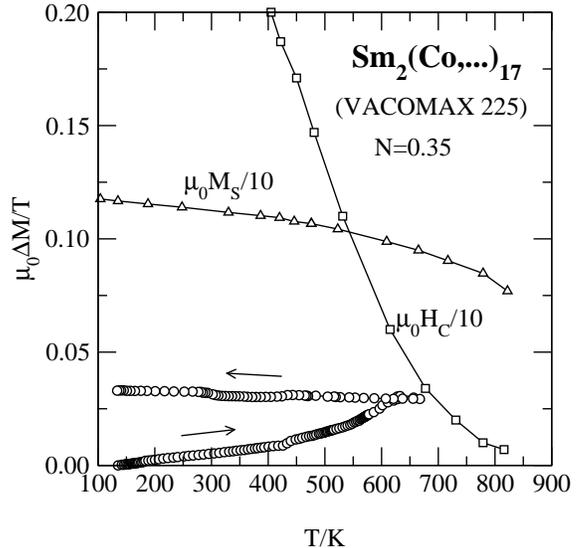}
\caption{The temperature dependence of the coercivity $H_C$, saturation
magnetization $M_S$ and the TR for a $Sm_2Co_{17}$ sample.}
\label{sm2co17tr}
\end{figure}
Nevertheless a small TR is present also in
$Sm_2Co_{17}$, as it is visible in Fig. \ref{sm2co17tr}. With regard to
the smallness of the effect, which amounts to not more than 3\%,
it is not yet clear, whether the effect is characteristic for all
$Sm_2Co_{17}$ magnets or whether it is caused by imperfections of the
microstructure, especially by the presence of a small content of
of $SmCo_5$ phase, as was argued in Ref. \cite{Scholl87} in connection
with the interpretation of virgin curves.\\

The $NdFeB$-type magnets consist of the magnetic active phase $Nd_{2-x}Dy_xFe_{14}B$
surrounded by nonmagnetic
phases. This guarantees for magnetic
decoupling of neighboring grains.
\begin{figure}[htb]
\psfig{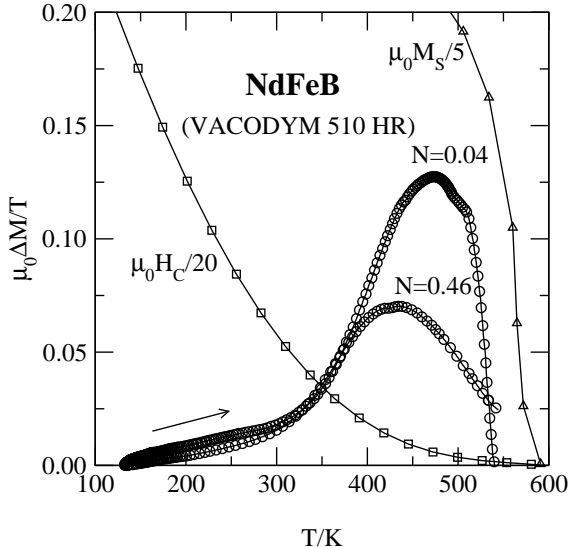}
\caption{The temperature dependence of the coercivity $H_C$, saturation
magnetization $M_S$ and the ITR of two NdFeB samples with N=0.04 and N=0.46 resp. \cite{Jahn02a}.}
\label{NdFeBTR}
\end{figure}
Within the main phase the nucleation
mechanism was shown to control magnetization reversal
\cite{Kronmueller91}. Apart from the much lower Curie temperature and
the spin reorientation below 137 K, the temperature
dependence of coercivity and saturation magnetization is
qualitatively the same as in $SmCo_5$ within the interesting
temperature region, see Fig. \ref{NdFeBTR}.
A remarkable
difference is that coercivity and magnetization vanish nearly at
the same temperature,
i.e. at \thc $\approx$ \Tc. The observed TR, as
depicted in Fig. \ref{NdFeBTR}, is  considerably smaller than in
$SmCo_5$, but, with 12 \% for the sample with very small
demagnetization factor it is doubtless that it is an intrinsic
characteristic of the magnet and can not be ascribed to
imperfections in the microstructure. Livingston reported TR effects of 3\%
for rod samples (N$\approx$0.17) with $H_C$ about 1 T/$\mu_0$ and 24 \% for a high coercive
sample (2.5 T/$\mu_0$) \cite{Livingston85}. Furthermore, the low coercive samples were
shown to contain a lot
of multidomain grains in the dc-demagnetized state, in contrast to the high coercive
specimen, which contained only a few. The same correlation between coercivity and TR was
also reported in \cite{Mueller88}.
\subsection{The ITR in Barium Ferrite}
\begin{figure}[htb]
\psfig{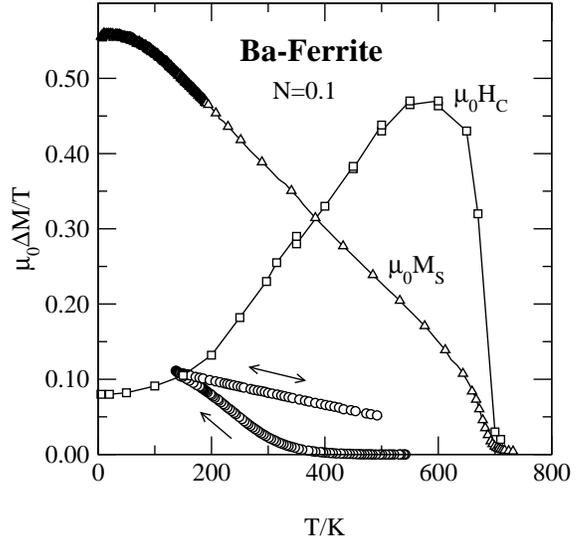}
\caption{The temperature dependence of the coercivity $H_C$, saturation
magnetization $M_S$ and the ITR of a barium ferrite sample
with N=0.1}
\label{BaFerriteTR}
\end{figure}

\begin{figure}[htb]
\psfig{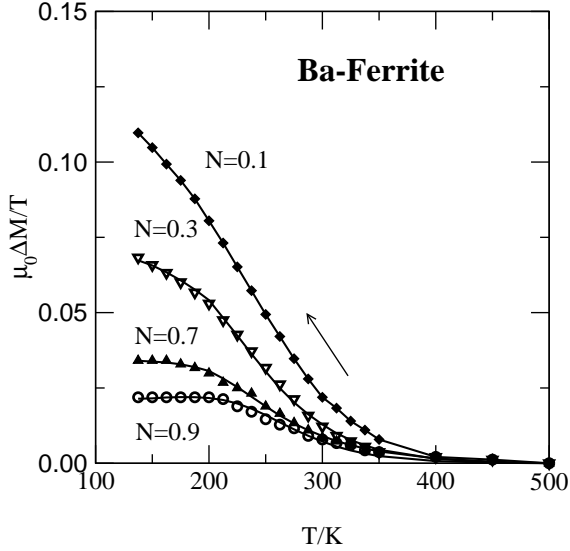}
\caption{The ITR for four
    samples with N=0.1, N=0.3, N=0.7, and N=0.9 respectively. The solid lines are
    theoretical fits.}
\label{BaFerriteN}
\end{figure}

In comparison with the above mentioned intermetallic magnets the ceramic
barium ferrite magnets exhibit a entirely different temperature dependence
of the coercivity, as shown in Fig. \ref{BaFerriteTR}.
That is why the normal TR cannot be measured in this substances.
Nevertheless an analogous
experiment, where an initially saturated sample is dc-demagnetized at
elevated temperature and afterwards cooled down, shows also a
remarkable remagnetization \cite{Jahn85}. This effect was called inverse thermal
remagnetization (ITR). As seen from Fig. \ref{BaFerriteN} we find again a
strong dependence on the demagnetization factor. Since the coercivity upon
cooling does not vanish one cannot observe a maximum. Instead one finds a
plateau at lower temperature. Barium ferrite magnets are single-phased
and the coercivity mechanism of barium ferrite is nucleation controlled.
In that respect, apart from the different temperature
dependence of the coercivity, the ITR of about 20\% is similar to the TR in $SmCo_5$-magnets.
\subsection{The field dependence}
\begin{figure}[!h]
\psfig{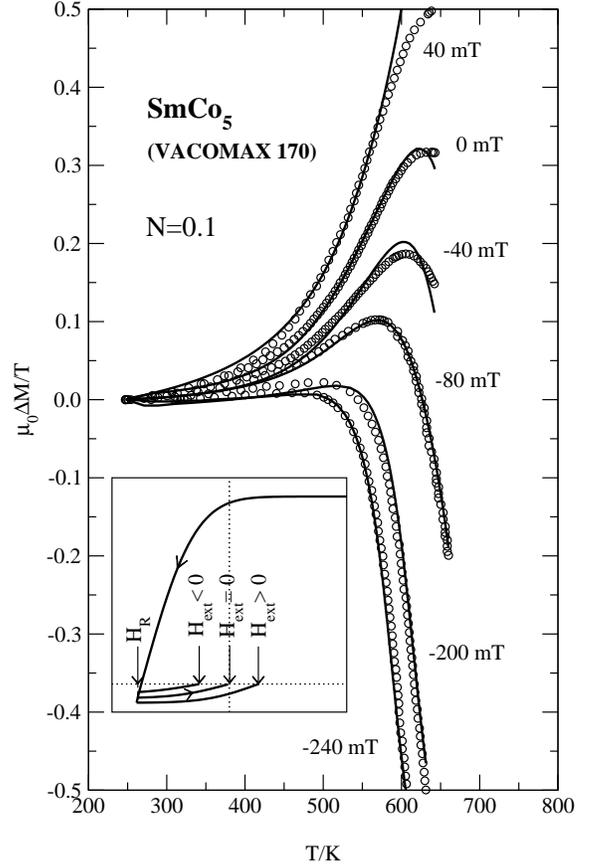}
\caption{Measured TR-curves $\Delta M(T)$ (circles) for
$SmCo_5$ \cite{Jahn02} for different external
steady fields applied during heating (small numbers
at the curves).
The solid lines are model calculations with parameters determined by
the $H_{ext}=0$-curve (cf. section 3). The inset shows schematically
how the initial state for the TR measurement is prepared.}
\label{SmCo5Feld}
\end{figure}

The dramatic dependence of the TR on the demagnetization factor shows that
a high sensitivity with regard to the internal magnetic fields is to expect.
The influence of a small superimposed steady field was studied in
\cite{Jahn85,Ivanov91,Jahn03}. To that behave the
standard experiment was modified in the following way:
After saturating the sample in a
field of (about 10 T for $SmCo_5$- and $Sm_2Co_{17}$-samples and about 2 T for the
barium ferrite samples) it was demagnetized by help of an opposite steady
field \hex{1}. If this field equals
the remanence coercivity $H_R$ the magnetization will go to zero after
switching off the field. Otherwise, if \hex{1} deviates a little bit from $H_R$,
the demagnetized state will be achieved on the recoil curve for
a small negative residual field $H_{ext}<0$, if \hex{1} is greater than $H_R$, and for
a small positive field otherwise, as the inset in Fig. \ref{SmCo5Feld} shows schematically.
The field $H_{ext}$ was then kept constant
while the sample was heated (or cooled in the case of barium ferrite) and the resulting
remanence enhancement $\Delta M(T)$ (TR-curve) was recorded.
A set of such measured TR-curves (points) is given in
Fig. \ref{SmCo5Feld} for SmCo$_5$ \cite{Jahn02}. The initial temperature
$T_0$ was 250 K. The solid lines are due to the theory, which will be explained below.
The cases with $H_{ext} = 0$, i.e. $H_1=-H_R$, correspond to a ``normal" TR-curve.
A systematic shift of the maximum temperature $T_{max}$, the maximum remanence enhancement
$\Delta M_{max}$ and also of the initial slope
$\Delta M_{TR}(T)/\Delta T$
\cite{Jahn02}
with increasing negative external field is observed.
\begin{figure}[htb]
\psfig{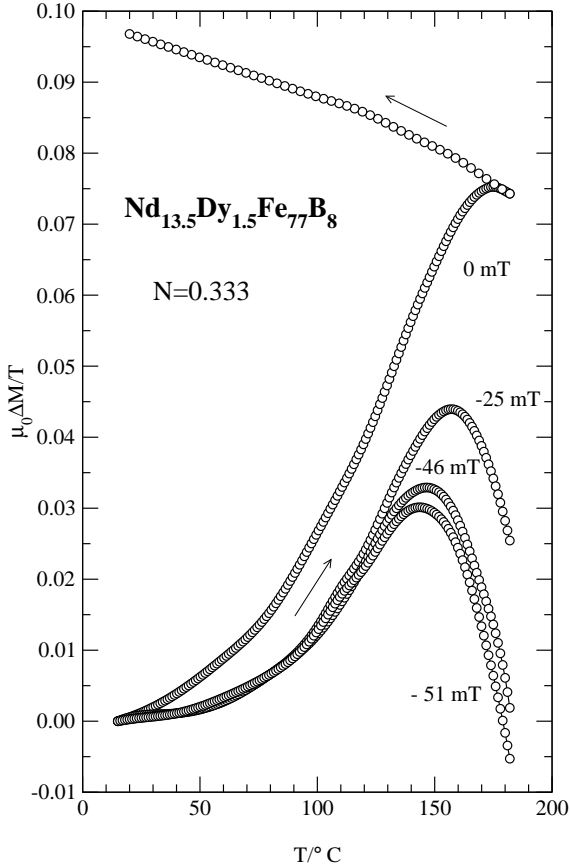}
\caption{Measured TR-curves $\Delta M(T)$ (circles) for
NdFeB  for different external
steady fields applied during heating (small numbers
at the curves) \cite{Ivanov91}.}
\label{NdFeBFeld}
\end{figure}
The same qualitative behavior can be measured in $NdFeB$-magnets (Fig. \ref{NdFeBFeld})
and $Sm_2Co_{17}$ magnets (Fig. \ref{Sm2Co17Feld}).
Besides the fact, that the absolute TR-change with field is smaller in $NdFeB$ and
even in $Sm_2Co_{17}$-samples, it looks very similar if the remanence-change with temperature
is related to the remanence enhancement of the standard experiment \cite{Ivanov91}.
\begin{figure}[htb]
\psfig{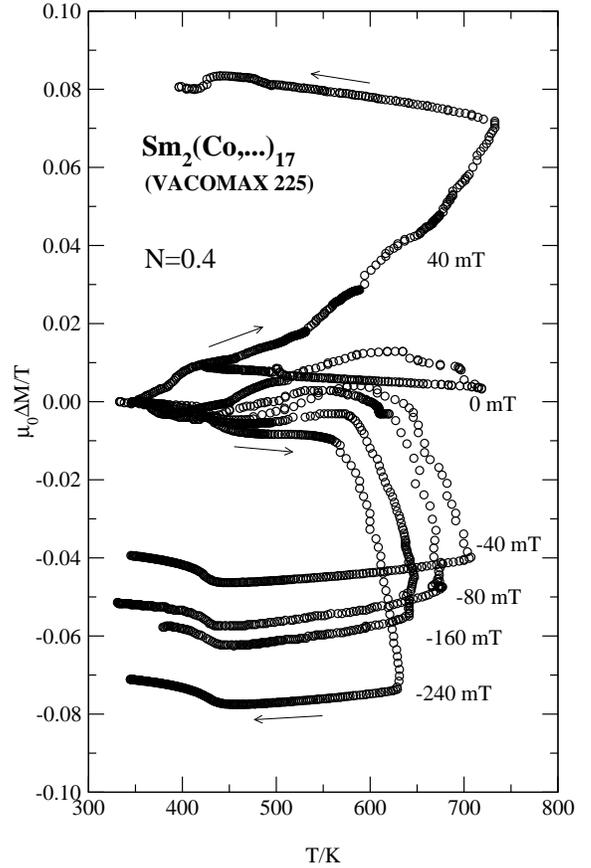}
\caption{Measured TR-curves $\Delta M(T)$ (circles) for
$Sm_2Co_{17}$ for different external
steady fields applied during heating and cooling resp. (small numbers
at the curves).}
\label{Sm2Co17Feld}
\end{figure}
Fig. \ref{BaFerriteFeld} shows the analogous experiment for the ITR measured
at a barium ferrite sample \cite{Schumann01}.
\begin{figure}[htb]
\psfig{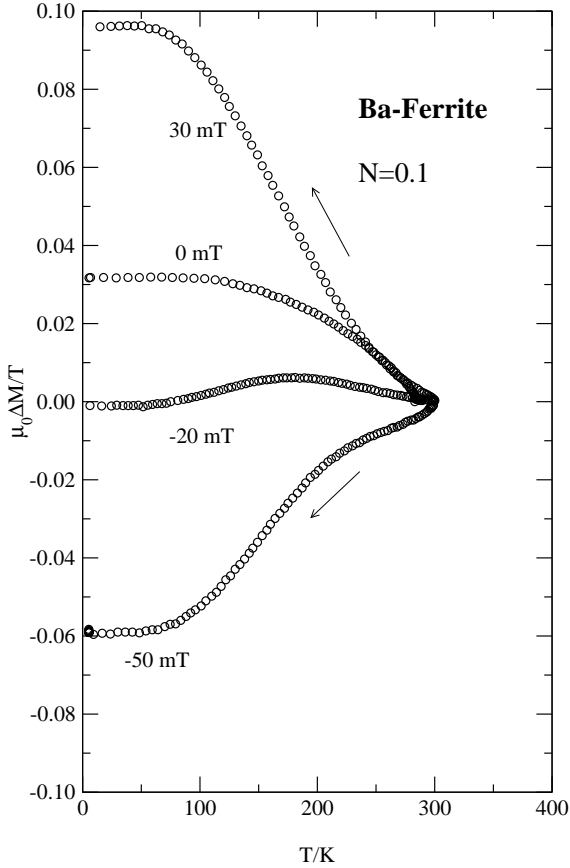}
\caption{Measured ITR-curves $\Delta M(T)$ (circles) for
barium ferrite for different external
steady fields applied during cooling (small numbers
at the curves)  \cite{Schumann01}.}
\label{BaFerriteFeld}
\end{figure}
As it is obvious from the Figs. \ref{SmCo5Feld}, \ref{Sm2Co17Feld},
(for $NdFeB$ cf. \cite{Jahn02}) a small positive field
increases the TR extremely. Otherwise
the TR may be suppressed at a certain negative field $H'$.
The order of magnitude of this "suppression field" $H'$ is to some extent a characteristic
of the different types of permanent magnets
\cite{Jahn85,Ivanov91}.
Fig. \ref{AlleTRmax} extracts the maximum TR values in dependence of $H_{ext}$
from Figs. \ref{SmCo5Feld}, \ref{NdFeBFeld}, \ref{Sm2Co17Feld}, and \ref{BaFerriteFeld}.
Comparing these results with other reported
measurements one finds $H'$ of the order of 200-500 mT for $SmCo_5$ \cite{Jahn85,Ivanov91,Jahn02} and
30-80 mT for $NdFeB$-magnets \cite{Jahn90,Jahn02}.
For $Sm_2(Co,\dots)_{17}$-magnets this value seems to be more
dependent on the chemical composition and technological treatment of the sample. Please note
the mT-scale in the lower picture of Fig. \ref{AlleTRmax} showing $\Delta M_{max}$ via $H_{ext}$
data for $Sm_{2}Co_{17}$.
Nevertheless $H'$ is in the same region as in $SmCo_5$. This may support the assumption
that the TR in $Sm_2Co_{17}$ is caused by a small amount
of $SmCo_5$ grains as was proposed in \cite{Scholl87}.
The suppression field for the ITR in the barium ferrite sample of Fig. \ref{BaFerriteFeld} is about 20 mT as
may be seen from the filled squares in Fig. \ref{AlleTRmax}.\\
\begin{figure}[hbt]
\psfig{file=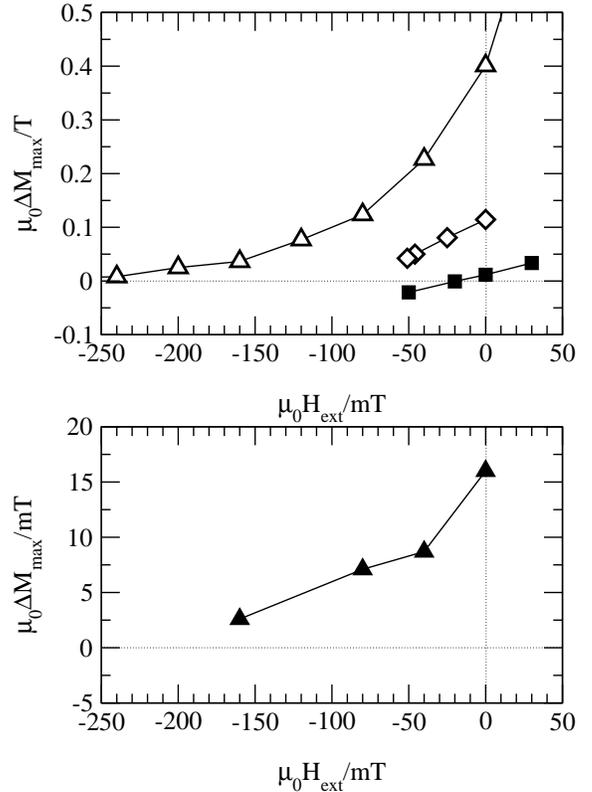,width=75mm}
\caption{The maximum TR-curves in dependence of the superimposed magnetic field for
the samples of Figs. \ref{SmCo5Feld}, \ref{NdFeBFeld}, \ref{Sm2Co17Feld}, \ref{BaFerriteFeld}.
Open triangles: $SmCo5$,
filled triangles: $Sm_2Co_{17}$,
diamonds: $NdFeB$,
squares: Barium Ferrite.}
\label{AlleTRmax}
\end{figure}
\subsection{Influence of the initial temperature T$_0$}
Ocularly the TR as well as the ITR are strongly connected with the temperature
dependence of the coercivity. The difference of the coercivity
$H_C(T_{max})-H_C(T_{0})$ is essentially for the TR. Thus
the remanence increase should be dependent on the initial temperature $T_0$ as well.
\begin{figure}[htb]
\psfig{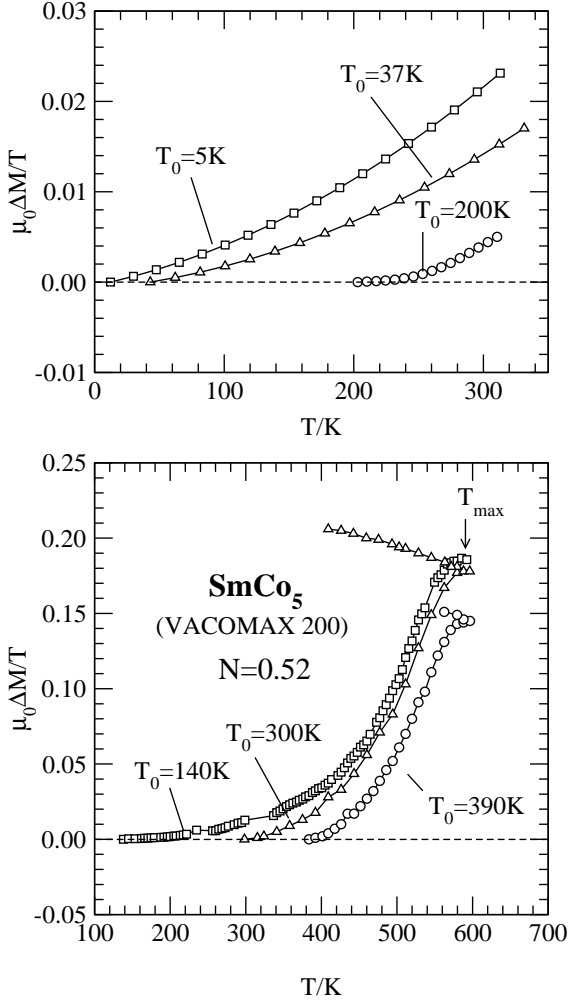}
\caption{  TR curves for different initial temperatures $T_0$. The upper picture
shows the initial TR for low initial temperatures.}
\label{SmCo5T0}
\end{figure}
In Fig. \ref{SmCo5T0}, we show the results for oblate discs cut from a well
aligned $SmCo_5$-magnet (VACOMAX200).
The upper part shows the initial irreversible TR at very low temperature,
i.e. in the high coercive region, in detail, whereas the lower part shows the
complete measurement to the maximum. Cooling down from the maximum the magnetization
increases reversibly - exactly proportional to
$M_S(T)$. More generally, if the TR experiment is interrupted at some temperature
$T_1$ the magnetization change while cooling follows exactly $m_1 \times M_S(T)$, with
$m_1$ being the ratio $\Delta M(T_1)/ M_S(T_1)$, see e.g. Fig. 1 in \cite{Ivanov91}.
On the one hand the amount of the TR is
considerably influenced by a change of the
initial temperature, on the other hand  only a small shift of the maximum
position to lower temperatures
is observed, if the initial temperature decreases.
%
\begin{figure}[htb]
\psfig{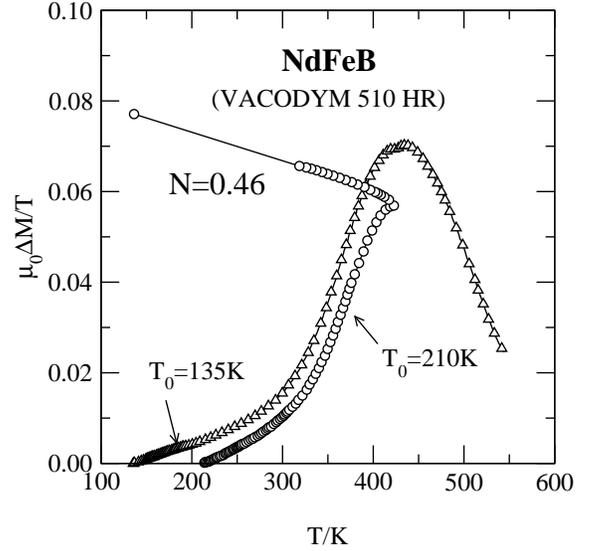}
\caption{TR curves for two different initial temperatures for a
NdFeB sample (NdFeB-HR 510) with N=0.46}
\label{NdFeBT0}
\end{figure}
%
A likewise behavior is observed in $NdFeB$ as depicted in Fig. \ref{NdFeBT0}.
Regarding
the ITR in barium ferrite, where the initial temperature has to be chosen higher
than the final temperature, we found again a strong dependence of the remanence
increase on the initial temperature, whereas the region where the curves run
into the plateau is slightly influenced only, as may be seen from
Fig. \ref{BaFerriteT0}.
\begin{figure}[htb]
\psfig{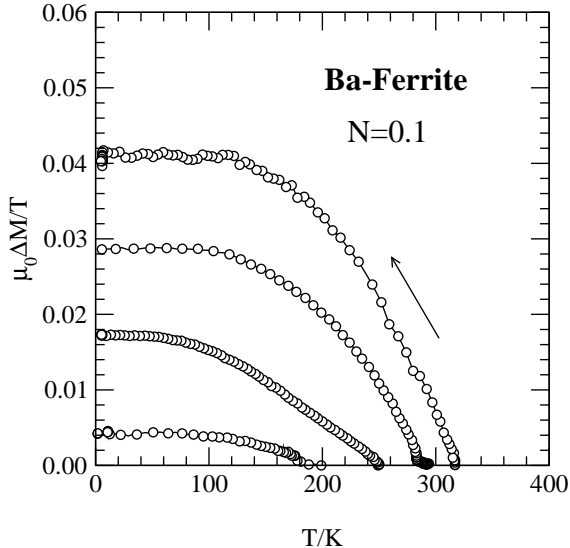}
\caption{ITR curves for three initial temperatures for a
isotropic sample.}
\label{BaFerriteT0}
\end{figure}
%
\subsection{Susceptibility and Repeating experiments}
Since heat treatment is a mayor step in permanent-magnet production
the possibility is to envisage that the sample after the first
TR experiment is irreversible changed. This is of course of minor
importance for the ceramic magnets but may be important in $SmCo_5$.
Since the maximum temperature in $SmCo_5$ is about 600 K, it is
low enough not to destroy the samples if the TR experiment does
not last to long. In \cite{Livingston84} the stability of the magnetization
of $SmCo_5$ against thermal cycling between 20 $^{\circ}$C and 75 $^{\circ}$C
was investigated. It was shown that it
needs a few cycles to stabilize the magnetization change. For a sample
which was partially dc-demagnetized (10\% lowered remanence),
it was shown that 16 h annealing at 75 $^{\circ}$C after ten cycles produces a
sharp positive $\Delta M$ jump while additional hold of 38 h after cycle 13 produced no further
significant change, besides a very small increase continuing up to 53 cycles.
In \cite{Jahn02} "repeating" experiments were reported, which differ a little
bit from the thermal cycling procedure. Contrary to \cite{Livingston84},
where after every TR cycle the sample was saturated with a high field,
the sample was dc-demagnetized  from the remanent state which it reached after a
cycle $T_0 \rightarrow T_{max} \rightarrow T_0$, i.e. without a new saturation.
The idea behind this experiment was to prove the assumption, that
the "hard grains", i.e. the fraction of grains which resist dc-demagnetization,
are not concerned by the TR experiment.
\begin{figure}[htb]
\psfig{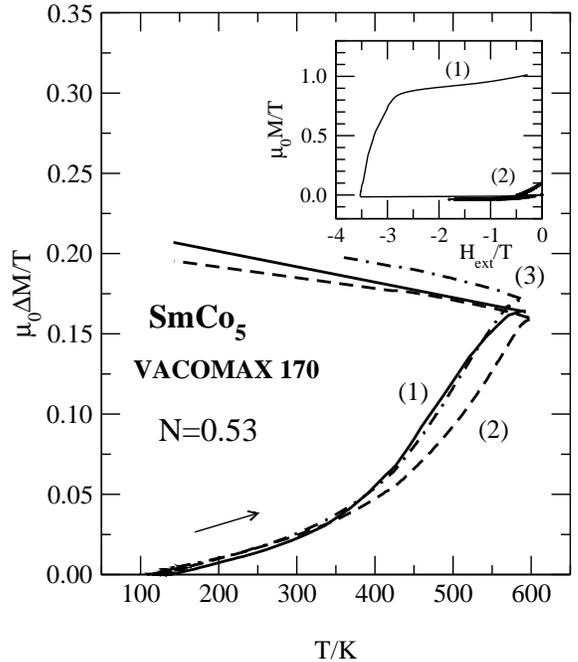}
\caption{TR curve (1) and two repetitions (2,3) for the same sample of
$SmCo_5$. The inset shows the dc-demagnetization curve for the first and second run.}
\label{SmCo5Repeat}
\end{figure}
Fig. \ref{SmCo5Repeat} shows a TR
experiment together with its two repetitions. The difference between the curves
are more likely due to the uncertainty in achievement of the dc-demagnetized
state. Thus it is merely the "weak" grain fraction, which is switched both
during dc-demagnetization and TR. After the first heating the sample is
demagnetized by a much lower field, as may be seen from the inset of Fig
\ref{SmCo5Repeat}, showing that the magnetic hardness of the
week grains was considerably lowered by heating to $T_{max}$. As a consequence
an increase of the susceptibility is observed.
\begin{figure}[htb]
\psfig{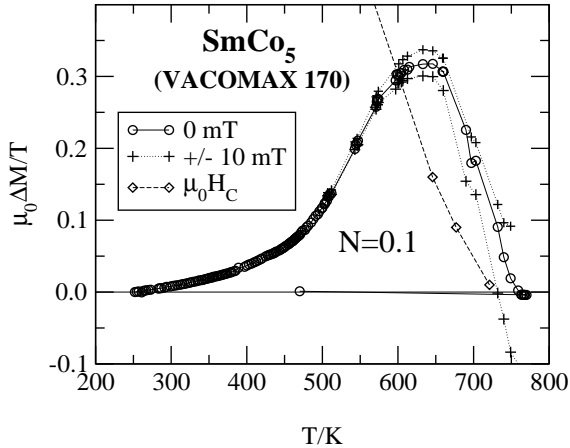}
\caption{ Measured TR curves for the same sample as in Fig. \ref{SmCo5Feld}
for $\mu_0 H_{ext} = \pm 10
\mbox{mT}$ ($+$).
The circles represent the values measured for $\mu_0 H_{ext} = 0$.
Furthermore  the $H_C(T)$ is indicated.}
\label{smco5plusminusexp}
\end{figure}
In the experiment shown in Fig. \ref{smco5plusminusexp} the TR curve was
started as normal in the dc-demagnetized state, but a very small field of 10 mT
was superimposed with alternating sign at every registration point of
the magnetization. This results in a "normal" and two slightly shifted TR
curves. The latter were taken as input for the calculation
of the susceptibility $\chi$, which was derived from the
un-sheared susceptibility $\chi'=\Delta M/\Delta H_{ext}$ according
to $\chi=\chi'/(1-N\chi')$ with $N$ being
the demagnetization factor of the sample.
%
\begin{figure}[hbt]
\psfig{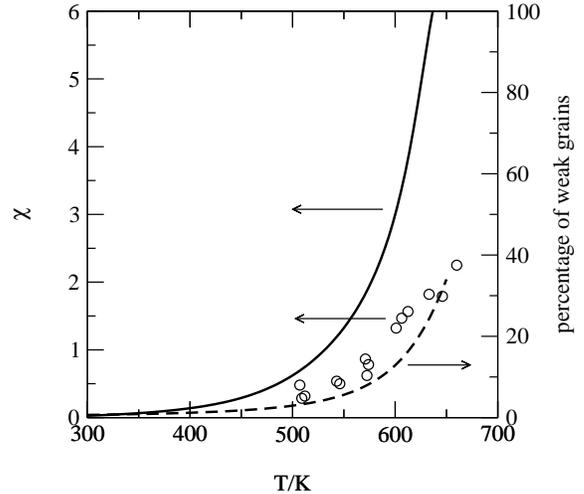}
\caption{Measured and calculated susceptibility (solid line)
$\mu_0H_{ext} = \pm 10 \mbox{mT}$ for the $mCo_5$-magnet from Fig. \ref{SmCo5Feld}.
Furthermore the theoretically calculated percentage of
``weak" grains is shown. }
\label{smco5susze}
\end{figure}
The plot in Fig. \ref{smco5susze}
shows a continuous uprise with increasing temperature, which becomes very large
beyond $T_{max}$. This sharp increase is due to the partial conversion of
the weak grain fraction into the MDS and due to the lowering of their switching
fields. This reduced coercivity of the weak grain fraction is frozen if the
sample is cooled back to $T_0$. Otherwise, the hard-grain fraction regains
nearly reversibly its initial state, what means that the interaction field
distribution due to the hard grains is-reestablished. In order to demagnetize
the sample another time one has to overcome the smaller coercivity of the weak-grain
fraction only. Afterwards the switched weak grains are located in nearly the same
environment as before. As a consequence  the TR is nearly the same in the repeating
experiments.
\subsection{Influence of texture.}
\begin{figure}[htb]
\psfig{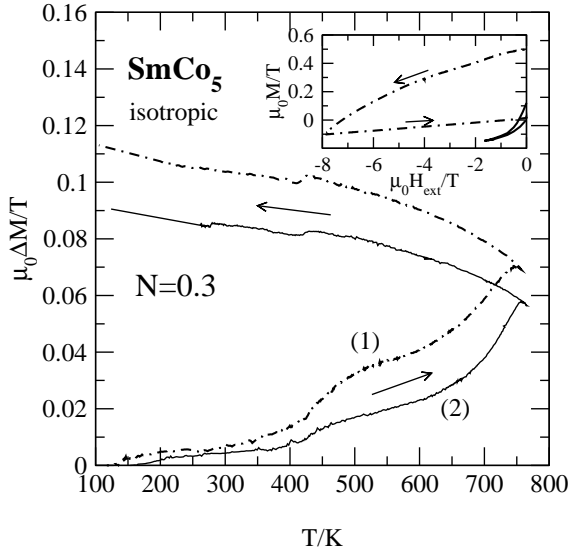}
\caption{TR curve (1) and one repetition (2) for an isotropic sintered sample of
$SmCo_5$ (Quality as VACOMAX 170, sintered and pressed without aligning field \cite{Rodewald}). The inset
shows the dc-demagnetization curve for the first and second run.}
\label{SmCo5RepeatIsotrop}
\end{figure}

Due to the technical importance, most of the reported TR experiments were done at
well-textured samples.
In Refs. \cite{Jahn02,Schumann01b} TR investigation at isotropic magnets were reported.
Comparing isotropic and well aligned samples of $SmCo_5$
one finds the same qualitative behaviour, but the TR of the isotropic specimen
is reduced roughly four times with respect to the well aligned sample. Furthermore the
difference of the TR curve and the first repetition differs a little bit more than
in the well aligned material,
as  may be seen from Fig. \ref{SmCo5RepeatIsotrop} in comparison with
Figs. \ref{SmCo5Kavalerova},\ref{SmCo5TR}, and \ref{SmCo5Repeat}.
Qualitatively the same behavior was reported for isotropic sintered barium ferrite \cite{Schumann01b}
Otherwise, an isotropic barium ferrite sample made from chemical precipitated
50 nm-powder, does not show any TR \cite{Jahn85,Seidel00}. To what extent this is
connected with texture is not completely clear. The nano-sized  grains are
nearly Stoner-Wohlfarth particles, with an entire different coercivity mechanism \cite{Stoner48}.
Furthermore the density is much lower than in the sintered barium ferrite samples,
what reduces the inter-grain interaction. That the latter plays an essential
role can also be concluded from the experiments of Lileev and Steiner
\cite{Lileev77}, who
measured the initial TR for $SmCo_5$ in the temperature region from 4.2 K to 270 K
for both well aligned sintered magnets and polymer-bounded samples with a $SmCo_5$
content of 20\% and 80\%.
They found a considerable reduction of the TR with decreasing $SmCo_5$
packing density.
\section{Theory}
\subsection{Early models}
During the TR experiments the macroscopic magnetic
energy is increased. This is only possible if the internal energy is
considerably decreased at the same time. During the initial dc-demagnetization
process a state is prepared where one half of the grains is upwards (i.e. in the
initial saturation direction) and the other half is downwards magnetized. Under
the assumption that the
intrinsic switching fields of the grains are not strong correlated with its position,
we will find a extreme inhomogeneous magnetization at a microscopic scale in the
dc-demagnetized state giving rise to considerable stray fields. While there is
agreement that it is this part of the energy which governs TR, two different
explanations for the mechanism behind TR were proposed. The first, introduced by
Livingston \cite{Livingston84}, assumes that the downwards magnetized grains,
which are in a single domain state before heating, decay into a multidomain
state upon heating. This model got support from the direct observation of
such grains in Kerr photographs taken from the surface of
dc-demagnetized $SmCo_5$-magnets \cite{Livingston84} and $NdFeB$-magnets \cite{Livingston85}.
Although these intriguing observations make clear that to some extent nucleation
of domain walls are involved in the TR, there are also
some arguments against the dominating role of the SDS-MDS mechanism.
Without any calculations
it is obvious that the maximum
TR effect may not be greater than 50\%, since after a complete decay of
the weaker half only the harder grains contribute to the magnetization.
This contradicts the observation of
a much greater effect measured for $SmCo_5$ in a closed circuit \cite{Kavalerova75}.
Furthermore, the number of the multidomain grains seen in the shown Kerr photographs
seems to be to low to explain the effect, despite the fact that it is
not completely clear to whether extent the surface observation is representative
for the bulk behavior.
The other model \cite{Jahn85} assumed that the grain
fraction with lower switching fields is first switched downward by the
dc-demagnetization procedure. After taking away the external field they
remain in a metastable state since their switching field is higher than
the interaction field caused by the surrounding upwards oriented, i.e. harder
magnetic, grains. Upon heating the switching fields are reduced and the weaker
grains are switched back in its initial direction. The decrease of the TR with
decreasing filling factor \cite{Lileev77} hints strongly at the
role of inter-grain interaction. Further support the interaction model got
from the ITR in barium ferrite, since contrary to
$SmCo_5$, $Sm_2Co_{17}$, and $NdFeB$, the
low temperature minimum of the coercivity is still high enough to prevent
a decay of grains into a multidomain state, thus the SDS-MDS mechanism can not
explain the ITR.
To clarify the role of multidomain grains Zaytzev et. al. \cite{Zaytzev88}
used Kerr micrography to count the number of multidomain grains in $SmCo_5$
magnets in the dc-demagnetized state at room temperature and at 120 $^{\circ}$C.
They demonstrated clearly that in the considered samples
three processes contribute to the TR: Downward magnetized grains
are either switched back or decay into the MDS under the influence of the
interaction fields due to surrounding grains and in MDS grains the upward
domains grow at cost of the downward domains.
Besides the fact that the investigated samples exhibit a lowered coercivity and
an unusual broad switching field distribution (SFD) in comparison with
usual commercial magnets and that surface effects may result in a slight
overestimation of the role of the MDS fraction from these experiments
it becomes clear that a theoretical model has to describe both the interaction
processes and the behavior of the MDS grains.
It is interesting that they also observed hard grains changing
from SDS to MDS.
Whereas the theory presented below uses a phenomenological ansatz, Lileev
et al. \cite{Lileev92} presented a model based on simulations
of interacting dipoles. This model, also taking into account the possibility
of MDS states, was originally developed for the simulation of minor loops of permanent
magnets \cite{Gabay91}. By taking into account the temperature dependence of the model
parameters it was also possible to simulate the TR. The shape of the only given curve
deviates far from experiments. The authors see the reason for that in
the mean-field treatment of the interaction contribution within their model.
Whereas Monte-Carlo
simulations allow to play with parameters it is almost impossible to determine
parameters by fitting to the experiment. For that purpose a
phenomenological description as given in the following seems more
appropriate.

\subsection{The inclusion model}
\begin{figure}[htb]
\psfig{file=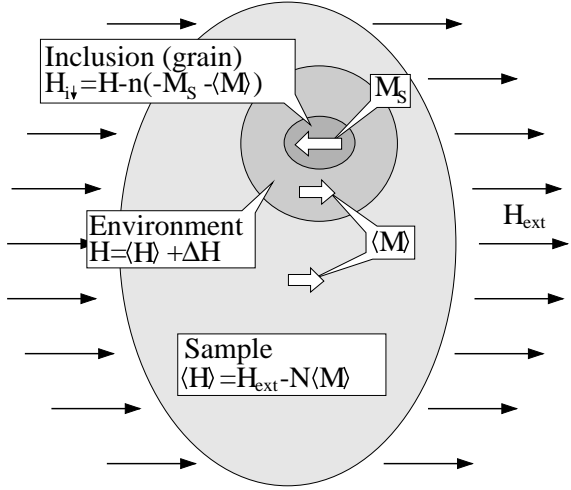,width=7.5cm}
\caption{ \label{cap:inclusion}
 The grain is considered as an inclusion with
           magnetization $\pm M_S$ (here we assumed $-M_S$, symbolized by the long white
           arrow) embedded in an environment differing in its magnetic field $H$ from the mean magnet
           field $\mean{H}$  within the sample due to the fluctuation $\Delta H$.
           Fluctuations of the magnetization in the environment are neglected, thus the
           magnetization in the environment is equal to the mean magnetization of
           the sample. This is symbolized by the two shorter white arrows of equal length.}
\label{inclusion}
\end{figure}

The outlined theory was developed
in Refs. \cite{Jahn85,Schumann87,Schumann01,Schumann01a}.
The poly-crystalline permanent magnet is an ensemble of high-coercive magnetic
uniaxial grains. To fix our model we have to define the properties of both a
single grain and
the ensemble. For the single grain we assume that the temperature dependent
saturation magnetization $M_S(T)$ should be known. Furthermore every grain
should be characterized by its temperature dependent switching field
$H_s(T)$, by its ``internal'' demagnetization factor $n$, by the direction of
the easy axis due to the high uniaxial anisotropy $\vec{c}$, and by its volume
$V$. Here $H_s(T)$ is the absolute value of the field needed for reversing the
magnetization of the grain in a closed circuit.
Regarding the ensemble we assume that
all grains exhibit the same $M_S(T)$, that the easy axis $\vec{c}$ of the
grains are completely aligned, i.e. we restrict the model to ideally textured,
single-phased magnets. These conditions are best fulfilled in $SmCo_5$ and
barium ferrite magnets, whereas $NdFeB$ magnets contain more than
one phase and $Sm_2Co_{17}$ magnets show a precipitation structure resulting in a different
switching behavior of a grain.
Furthermore we allow that the
grains differ in their switching fields $H_s$,
but the distribution of these switching fields should be known.
In non-interacting  ensembles the switching field distribution may be
determined from remanence measurements \cite{Chantrell86} , but
if the grains interact strongly this method fails.
In Ref. \cite{Jahn90} both the thermal remagnetization and
the  change of the $H_s$ spectrum with increasing temperature of a $SmCo_5$ sample
was measured,
whereby the latter was approximately determined from $dM/dH$. This is justified
if reversible magnetization processes are negligible as it is the case in $SmCo_5$-magnets,
For simplicity we assume a Gaussian normalized with respect
to the region $0<H_s<\infty$ \cite{Schumann87},
\beq
\romg(H_s)&=&\norms \,\, {\rm e}^{
-\left (\frac{H_s-\bar{H}_s}{\sigmas} \right)^2}
,
\label{gs}
\eeq
with the ``mean switching field'' $\bar{H}_s$ and the distribution width $\sigmas$.
The temperature dependence (not the absolute value!) of the switching
fields of all grains are equal.
For the
moment we have to regard $\bar{H}_s$ and $\sigmas$ as model parameters. Fortunately
we will see later on that the theory delivers a possibility to relate
$\bar{H}_s$ to the coercivity $H_C$ of the sample.
An central point of TR is the treatment of the grain-grain interactions resulting in
a collective behavior of adjacent grains.
On a macroscopic length scale the internal magnetic fields and the magnetization of the
sample are homogeneous.
But, if  the length scale is reduced to the order of some grain diameters the
magnetization becomes coarser and the inhomogeneities gain influence.
Averaging over volume elements containing a few grains only, yields values,
which deviate from the average.
Thus, we consider every grain as an inclusion embedded in a local environment,
which may differ stochastically in its magnetic field and magnetization
from the related averaged values. We show this schematically in
Fig. \ref{inclusion}.
For simplicity we approximate the magnetization in the vicinity of the
inclusion by the mean magnetization
$\mean{M}$, neglecting the mentioned magnetization fluctuations.
Otherwise we allow for field fluctuations  $\Delta H = H-\mean{H}$
around the mean internal field
in the environment of a grain. The mean internal field is related to
the external applied field $\Hex$ according to
\beq
\mu_0\mean{H}=\mu_0\Hex-N\mean{M} \komma
\label{extdemag}
\eeq
with $N$ being the demagnetization factor of the sample.
The field fluctuations are assumed to be Gauss distributed
\cite{Schumann87,Mueller87},
\beq
\romf(H)&=& \frac{1}{\sqrt{\pi} \sigmafhat} \,\,
{\rm e}^{
-\left(\frac{\Delta H}{\sigmafhat}\right)^2}
\\ \nonumber &&
\qquad \mbox{with} \qquad \Delta H
= H-\mean{H} \punkt
\label{fh}
\eeq
what is a consequence of the central limit theorem of probability
theory. Furthermore, the fluctuation width $\sigmafhat$ has to be
an even function of the magnetization itself.
\beq
\sigmafhat &=& \sigmaf \,\xi(m) \quadbox{with} m=\frac{\mean{M}}{M_S} \punkt
\label{sigmafvonM}
\eeq
This becomes obvious since
in the saturated magnetized state the fluctuation width is zero (remember:
ideal texture, all grains the same phase). Otherwise it is
maximum for the dc-demagnetized state with random distribution of upward
and downward magnetized grains. A Taylor expansion of $\sigmafhat(\mmm{})$ to second order
which fulfils these two boundary conditions yields
\beq
\xi(m) &=&  ( 1 - m^2 ) \punkt
\label{xivonM}
\eeq
In general $\xi(m)$ depends on the correlation of neighboring
grains. For instance the field of a simple cubic array of uncorrelated dipoles, randomly
oriented in $\pm$z-direction, yields
\beq
\xi(m) &=&  \sqrt{ 1 - m^2} \punkt
\label{sigmafvonMsqrt}
\eeq
Of course the local fields fluctuate around the mean internal field
also in direction, thus, strictly
speaking eq. (\ref{fh}) accounts for the fluctuations of the z-component only.
Furthermore the deviation of the field direction from the easy axis gives
rise to rotation processes, but this
is surely a small effect in ideally aligned magnets with high anisotropy
constants.
In the early works on the TR theory a further point was stressed \cite{Jahn85,Schumann87,Lileev92}:
The field dependence of the switching field. This is indeed a crucial point, if one attempts to describe
repeating experiments or the virgin curve, but can be neglected in the first TR experiment, if
$H_R$ is large enough to make most of the down-switched particles hard, as it seems to be the case
in $SmCo_5$ and barium ferrite. Taking into account this process is in principle not
difficult, since it simply scales the SFD of the switched particles. This would
introduce at least one additional parameter. We abstained from doing so by restricting the
theory to the first TR run only.
\subsection{Calculation of the Isothermal Demagnetization Curve}
\begin{figure}[htb]
\psfig{file=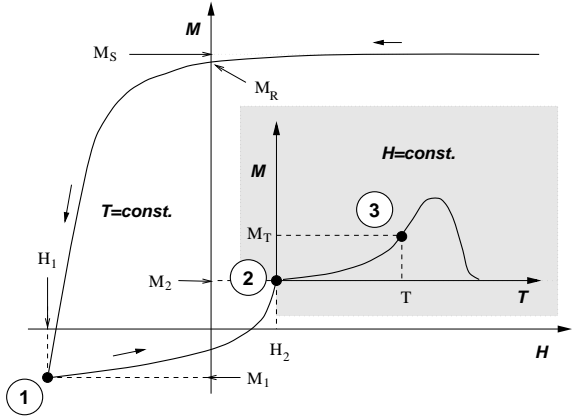,width=75mm}
\caption{Generalized scheme of the a TR- Experiment:
After saturation the sample is
demagnetized isothermally by a steady field to an arbitrary point of the
demagnetization curve, indicated by $H_1$ and $M_1$. Afterwards the field
is reduced to an arbitrary value $H_2$. The related magnetization is $M_2$.
Finally the sample is heated (TR-Experiment) or cooled (ITR-Experiment)
with the external magnetic field kept constant.}
\label{prinzipH1H2}
\end{figure}

The first step is the calculation of the demagnetization curve from the
saturated state down to the magnetic field $\hex{1}$ in the third quadrant.
see Fig. \ref{prinzip}. To save the opportunity to describe also the influence of a small
steady field or a non-zero initial remanence on the TR we use a more general scheme of
the TR experiment as shown in Fig. \ref{prinzipH1H2}.
After the saturation all grains are up ($\uparrow$) magnetized.
They posses
their maximum switching fields and $\mmm{}=M_S$ holds.
If the external field is lowered to a value $\hex{1}$ (c.f. Fig. \ref{prinzipH1H2}
the magnetization of a grain is switched down ($\downarrow$) if its internal
field $\hiup{1}$ is lower than $-H_s$ and the internal field
after switching $\hidown{1}$ will be smaller than $+H_s$.
Thus we have the two conditions
\beq
\hiup{1}&<&-H_s \quad \quad \mbox{and}\quad \quad  \label{schalt1} \hidown{1}<+H_s  \quad .
\eeq
The internal fields before and after switching are with respect to the above
introduced approximations
\begin{eqnarray}
\hiup{1} &=& H - n ( M_s -\mmm{1})
\eeq
\mbox{and}
\beq
\hidown{1} &=& H - n( - M_s-\mmm{1}) \punkt
\end{eqnarray}
Here $H$ is the local field in the environment of the grain.
A variation of the external field $\Hex$ results in a change of $\mmm{}$
on the one hand and on the other hand it changes the probability $f(H)dH$
that the field in the environment of a grain is between $H$ and $H+dH$ due
to the relation (\ref{extdemag}) and eq. (\ref{fh}).
If the first condition from eq. (\ref{schalt1}) is fulfilled but not the second,
the grain cannot jump into a stable state.
If the grain is large enough
it solves the conflict by an incomplete jump, what turns its single-domain
state (SDS) into a multi-domain-state (MDS).
We will call such grains ``weak''.
Whereas the ``hard'' grains can exist in states with $\pm M_S$ only, the
magnetization
of a weak grain $\mmm{i}$ is an average over the upwards and downwards
magnetized volume fractions within the grain.
From phase theory of
180$^{\circ}$-domains we get for the averaged magnetization of the
i$\rm ^{th}$ grain in dependence on the local field in the environment
\beq
\mmm{i}&=&\mmm{} + \frac{H}{n} \quad .
\label{weakmagnetization}
\eeq
Opposite to the hard grains the weak grains have no memory since the
magnetization follows the local field immediately as long as $|\mmm{i}|<M_S$
holds. Otherwise they are saturated up- or downwards.
If one calculates the probability that a grain with a given $H_s$ is in a MDS
the switching condition (\ref{schalt1}) limits the integration over the
local field distribution from above,
whereas the back-switching condition $\hidown{1}>+H_s$ limits from below. The
related probability is
\beq
p_{w}(H_s)&=& \int\limits_{H_L}^{H_H}\!\!dH\; {\rm f}(H)
\eeq with \beq    
H_H&=&-H_s+nM_S-n\mmm{} \\
H_L&=&H_s-nM_S-n\mmm{}
\punkt
\eeq
The requirement, that the upper limit has to be greater then
the lower one, yields the condition
$H_s<n M_s$.
Thus, grains with switching fields smaller than their own internal
demagnetizing field are weak.
The magnetization of the volume fraction of the weak grains with a given $H_s$
is
\beq
\mmm{w}(H_s) &=&
M_S \int\limits_{H_H}^{\infty}dH\,  \romf(H)
\nonumber \\ &&
-M_S \int\limits_{-\infty}^{H_L}dH\,  \romf(H)
\nonumber \\ &&
+\int\limits_{H_L}^{H_H}dH\, \romf(H) \, \mmm{i}
\label{mmmwhs}
\punkt
\eeq
The integral may be easy evaluated, where we have to take for
the magnetic field $\hex{1}$ and the related magnetization $\mmm{1}$ at
the hysteresis curve.
For hard grains $H_L$ is higher than $H_H$. If these grains fulfill
the switching condition, the second condition on eq (\ref{schalt1})
will be fulfilled automatically.
The probability to find a grain downwards magnetized if a field \hex{1}
is applied is
\begin{eqnarray}
p_1(H_s)&=&\int\limits_{-\infty}^{+\infty}\!\!dH\; {\rm f}_{(1)}(H) \;
\times \nonumber \\ && \!\!\!\!\!\!\!\!\!\!\!\! \times
\Theta(-H_s-\hiup{1})\;
\Theta(H_s-\hidown{1}) \, ,
\label{p1def}
\end{eqnarray}
with $\hiup{1}$ and $\hidown{1}$ being the fields within the grain in the
upwards and downwards magnetized state resp. and
$\Theta(x)$ is the Heaviside function. ${\rm f}_{(1)}$ is the field
distribution eq. (\ref{fh}) with $\mean{H}=\hex{1}-N\mmm{1}$.
After carrying out the integration in eq. (\ref{p1def}) we get
for the magnetization $\mmm{1}=M_S\, (1-2\,p_1(H_s))$ of a fraction
of hard grains with a given $H_s$
\beq
\mmm{1}(H_s)&=& - M_S \,\romerf(x_H)
\eeq with \beq
x_H&=&
\\&&
\mbox{\hspace{-1cm}}
\frac{-H_s+n M_s -n \mmm{1}-\hex{1}+N \mmm{1}}{\sigmafhat(m_1)}
\punkt
\nonumber
\eeq
The total magnetization  of the sample one gets from averaging with
respect to $H_s$
\beq
\mmm{1}&=&\int\limits_{0}^{nM_S}\,d\,H_s\,\romg(H_s)\, \mmm{w,1}(H_s)
\nonumber \\ &&
\!\!\!\!\!\!\!
 +
\int\limits_{nM_S}^{\infty}\,d\,H_s\,\romg(H_s)\, \mmm{1}(H_s)
\label{demagw}
\eeq
Here the integration has been split due to the different contributions of
the weak and hard grains. For the weak grains we have to use
eq.  (\ref{mmmwhs}).
Since the right hand side of eq. (\ref{demagw}) depends on the mean
magnetization $\mmm{1}$ itself, we have to solve this implicit equation
numerically.
\subsection{ Calculation of the Isothermal Recoil Curve}
Next, the external field is changed from
$\hex{1}$ to $\hex{2}$ with $\hex{2}>\hex{1}$ via the recoil curve.
Some of the hard ``down''-grains are switched back,
whereas the weak grains shift their magnetization reversible.
The related mean magnetization is $\mmm{2}$.
The contribution of the weak grains results from eq. (\ref{mmmwhs}), if
we insert now $\hex{2}$ for $\Hex$ and $\mmm{2}$ for $\mmm{}$ resp. followed by
an integration from zero to $nM_S$ with respect to $H_s$.
The contribution of the hard
grains is more difficult to calculate, due to
the memory effect. Let us consider the probability $\wdu(H_s)$ that a hard
grain with a given $H_s$ was
switched by $\hex{1}$ and switched back by $\hex{2}$ afterwards.
To calculate this probability it is inevitable
to make an assumption on the correlation between the fluctuation of
the local field $H_1$ which acts on a grain if $\hex{1}$ is applied and the
fluctuation of $H_2$ according to the external field $\hex{2}$. Of course, if
$\hex{2}$ is only slightly different from $\hex{1}$ it is unlikely that the
neighborhood of a grain changes considerably, so that $H_{1}$ and $H_{2}$
should be strongly correlated. With increasing distance between
$\hex{1}$ and $\hex{2}$ this correlation will vanish, due to the multitude of
switching processes, hence we can average with
respect to $H_1$ and $H_2$ independently, if the difference $\hex{2}-\hex{1}$
is not to small. Thus the probability $\wdu$ decouples accordingly
$ w_{\downarrow \uparrow}(H_s)= p_1(H_s) q_2(H_s)$ with $p_1(H_s)$ from
eq. (\ref{p1def})
and
\beq
q_{2}(H_s)&=& \int\limits_{-\infty}^{\infty} dH \, \romf_{(2)}(H)\,
\times \nonumber \\ && \times \,
                  \Theta(\hidown{2}-H_s) \, \Theta(\hiup{2}+H_s)
\nonumber \\
                  & =& \frac{1}{2} \, \Big ( 1-\romerf(y_L) \Big )
\eeq
with
\beq
y_L &=&
\\&&
\mbox{\hspace{-1cm}}
\frac{H_s - n M_s - n \mean{M}_2 - \hex{2}
+ N \mean{M}_2}{\sigmafhat(m)} \punkt
\nonumber
\eeq
Due to the mentioned magnetization changes in the environment
of a grain it may happen that grains which resisted $\hex{1}$ are switched
down by $\hex{2}$ (``up''-``down''-contribution). The related probability
factors also
\beq
\wud(H_s)&=& \Big ( 1-p_1(H_s) \Big ) \, p_2(H_s)\eeq
with $p_1(H_s)$ again from eq. (\ref{p1def}) and
\beq
p_{2}(H_s)&=& \int\limits_{-\infty}^{\infty} dH \,\romf_{(2)}(H)
\times \nonumber \\ &&  \times \,
                        \Theta(-H_s-\hiup{2})\,\Theta(H_s-\hidown{2})
 \nonumber \\
          &=& \frac{1}{2} \, \Big ( 1+\romerf(z_H) \Big )
\eeq
with
\beq
z_H &=&
\\&&
\mbox{\hspace{-1cm}}
\frac{- H_s + n M_s - n \mean{M}_2 - \hex{2}
+ N \mean{M}_2}{\sigmafhat(m)}
\punkt
\nonumber
\eeq
There are two further probabilities corresponding to the remaining two
histories, which a grain may experience, i.e.
the probability that it resisted both $\hex{1}$ and $\hex{2}$ (
``up''-``up'' contribution) and the probability that a grain was switched
by $\hex{1}$ but resisted back-switching by $\hex{2}$
(``down''-``down'' contribution). We find
\beq
\wuu(H_s)&=& \Big ( 1-p_1(H_s)\Big ) \, \Big ( 1-p_2(H_s) \Big ) \label{wuu}
\eeq and \beq
\wdd(H_s)&=& p_1(H_s) \, \Big( 1-q_2(H_s) \Big ) \label{wdd}
\punkt
\eeq
Thus hard grains with a given $H_s$ contribute to the magnetization
\beq
M_2(H_s)&=& M_s \Big ( \wuu(H_s) - \wud(H_s)
\nonumber \\ && \quad
                     + \wdu(H_s) - \wuu(H_s) \Big )
\punkt
\eeq
Averaging with respect to $H_s$ yields the recoil curve
for the mean magnetization $\mmm{2}$ in dependence on
both magnetic fields $\hex{1}$ and $\hex{2}$:
\beq
\mmm{2}&=&\int\limits_{0}^{nM_S}\,d\,H_s\,\romg(H_s)\, \mmm{w,2}(H_s)
\\&& \nonumber
+
\int\limits_{nM_S}^{\infty}\,d\,H_s\,\romg(H_s)\, \mmm{2}(H_s)
\label{recoilw}
\punkt
\eeq
%
\subsection{Calculation of the (Inverse) Thermal Remagnetization}
The above calculation of the demagnetization and
the recoil curves was done with the implicit understanding
that the magnetization changes
isothermally at a temperature $T_0$.
The actual TR (ITR) occurs if
a dc-demagnetized magnet is heated (cooled) while $\hex{2}$ is kept constant.
Changing the temperature
has two effects. On the one hand the saturation magnetization decreases
with  increasing temperature and on the other hand the switching fields $H_s$
are changed. Whereas the temperature dependence  of the saturation
magnetization can be measured easily, it is impossible to measure it for
the switching fields. Fortunately, one can solve
eq. (\ref{mmmwhs}) for $\bar{H}_s$ with $\hex{1}=H_C$ and $\mean{M}_1=0$ \cite{Schumann01}.
$H_C(T)$ and $M_S(T)$ are taken from measurements .
If the influence of the weak grains is negligible, $\bar{H}_s(T)$
is related to $H_C(T)$ and $M_S(T)$ by the simple formula
$ \mu_0\bar{H}_s(T)=\mu_0H_C(T)+nM_S(T) $ \cite{Schumann01}.
Furthermore we adopt the temperature dependence of $\bar{H}_s$ for arbitrary
values of the switching fields, i.e
\beq
H_s(T)=H_s(T_0)\times(\bar{H}_s(T)/\bar{H}_s(T_0)) \punkt
\eeq
The calculation of the magnetization
$\mean{M}_2^T$ at an enhanced temperature $T$ is similar to the calculation
of the magnetization $\mean{M}_2$ by help of eq. (\ref{recoilw}),
whereby the temperature induced deformation of the switching field distribution
has to be regarded.
Due to the assumption on $H_s(T)$ and the normalization of the switching
field distribution  follows that $\sigmas$ has the same temperature dependence
like $H_s(T)$.
The magnetization of a weak grain with
given $H_s$ at the temperature T results from eq. (\ref{mmmwhs}) with
the appropriate $H_s(T)$, $M_S(T)$, and $\mean{M}_{2}^T$ inserted.
For the hard-grain contribution we have to calculate the probabilities
$\wuu$, $\wud$, $\wdu$, and $\wdd$,
however for the increased (lowered in case of ITR) temperature T.
We find for the magnetization of the hard-grain fraction with given $H_s$
\beq
M_2^T(H_s)&=& M_S^T  \Big ( \wuu^T(H_s) - \wud^T(H_s)
\nonumber \\&&
 +  \wdu^T(H_s) - \wdd^T(H_s) \Big )
 \punkt
\eeq
It is obvious that the probability
$p_1(H_s)$ remains unchanged, since the grains were switched downwards
at the initial temperature $T_0$. Therefore we have
\beq
\wuu^T(H_s)&=&\!\!\!\!\!\!
\Big ( 1-p_1(H_s) \Big )\,
                      \Big ( 1-p_2^T(H_s) \Big ) \label{wuut}\\
\wud^T(H_s)&=& \Big ( 1-p_1(H_s) \Big ) \, p_2^T(H_s)\komma\label{wudt}\\\
\wdu^T(H_s)&=& p_1(H_s) \, q_2^T(H_s)\komma\label{wdut}\\\
\wdd^T(H_s)&=& p_1(H_s) \, \Big ( 1-q_2^T(H_s) \Big ) \punkt
\label{wddt}
\eeq
The probabilities
$q_{2}^T(H_s)$ and $p_{2}^T(H_s)$
become temperature dependent both directly due to $H_s(T)$, $M_s(T)$,
and $\sigmaf(T)$ and indirectly via $\mean{M}_2^T$.
The temperature dependence of $\sigmaf(T)$ was assumed to be that of $M_S$,
since the fluctuations are caused by the inhomogeneities of the
magnetization. Finally the TR or the ITR
can be calculated from the following implicit self-consistent equation
\beq
\mean{M}_2^T&=&\int\limits_{0}^{nM_S}\,d\,H_s\,\romg(H_s)\, \mmm{w,2}^T(H_s)
\nonumber \\&&
+
\int\limits_{nM_S}^{\infty}\,d\,H_s\,\romg(H_s)\, \mmm{2}^T(H_s)
\label{remag}
\punkt
\eeq
\subsection{Influence of approximations}
The theory teaches us, which processes are necessary to describe different parts of the
TR curve qualitatively right.
In order to describe a residual remanence beyond \thc,  $H_S(T)$ has to be determined
self-consistently from
$H_C(T)$ and $M_S(T)$.
\begin{figure}[!h]
\psfig{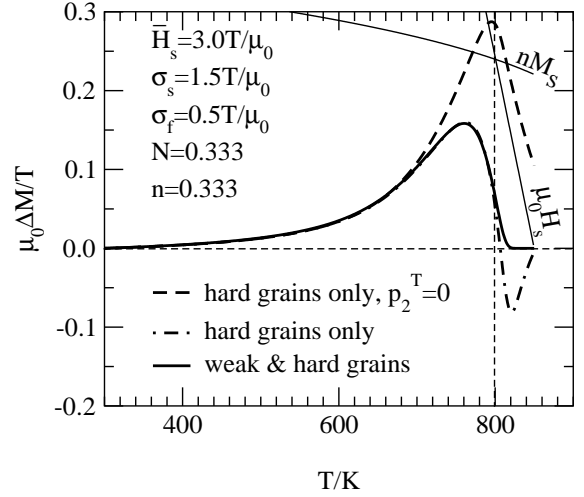}
\caption{
\label{abb:approximations} The TR for different levels of
approximation. The solid line gives the results of the presented theory.
The dot-dashed line
results from neglecting the weak grain fraction and for the calculation
of the dashed line, both the weak grains and $p_2^T$, i.e. the probability,
that a grain which resisted $\hex{1}$ will be switched while heating at
$\hex{2}=0$, were neglected.
}
\label{approximations}
\end{figure}
In Fig. \ref{approximations} we study different approximation levels.
The influence of weak grains
is small as long as the temperature is below $T_{max}$. For $T>T_{max}$
the neglect of the weak grains would result in a
qualitative different behaviour, since we get a sign change of the TR
before it finally vanishes, as shown in Fig. \ref{approximations} in
dashed-dotted manner.
The reason for that peculiarity is made clear in Fig.
\ref{probabilities}, where we have plotted the temperature dependence of the
probabilities $\wuu$, $\wud$, $\wdu$, and $\wdd$ regarding the hard grain
fraction.
It is obvious that at first
$\wdu$ increases and when it drops down $\wud$ rises up, thus
explaining the sign change. This behaviour is not observed if the probability
$p_2$, i.e. that a grain which resisted $\hex{1}$ will be switched by $\hex{2}$,
is neglected, as it was done in former theories, but then an overestimation of
the TR occurs, as visible in Fig. \ref{approximations}. For the description
of the TR at higher temperatures it is necessary to take into account,
that above $T_{max}$ even the switching fields of the hardest grains are
lowered enough to be switched by the still existing stray fields. The latter
do not vanish until the magnetization breaks down.
\begin{figure}[!h]
\psfig{file=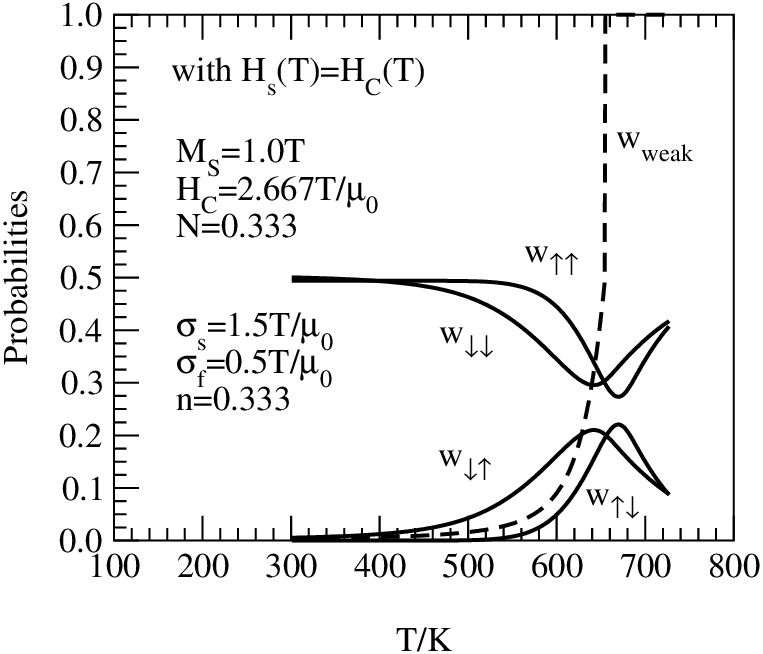,width=7.5cm}
\caption{
\label{abb:probabilities} The probabilities $\wuu^T$,
$\wud^T$, $\wdu^T$, and $\wdd^T$ (cf. eqs. (\ref{wuut}-\ref{wddt}))
in dependence on the temperature. The dashed line shows the
ratio of the volume of weak grains to the total volume.
\label{probabilities}
}
\end{figure}
It is worth to point out, that weak grains are not necessary to get effects of
more than 50 \%.
Such a large TR one gets due to the feedback of the magnetization via the
internal demagnetizing fields. In some sense this models
the well known avalanche effects observed during hysteresis measurement,
which should also occur to a certain extent if the sample remagnetizes.\\

In the early works on the TR a further point was stressed \cite{Jahn85,Schumann87,Lileev92}:
The field dependence of switching field. This is a crucial point, if one attempts to describe
repeating experiments or the virgin curve, but can be neglected in the first TR experiment, if
$\hex{1}$ is large enough to make most of the down-switched particles hard, as it seems to be the case
in $SmCo_5$ and barium ferrite. Taking into account this process is in principle not
diffult, since it simply scales the SFD of the switched particles. This would
introduce at least one additional parameter. We abstained from doing so by restricting the
theory to the first TR run only.
Taking into account all probabilities and weak grains
the model describes the TR dependence on the demagnetization factor $N$ of
the sample and the initial temperature $T_0$ well,
as was shown in Ref. \cite{Schumann01} for the TR and in Ref. \cite{Schumann01a} for the ITR.
In the following we concentrate on the investigation of the influence of internal model parameters.
\subsection{Comparison to the experiment}
The limitations of the presented model, i.e. the assumed perfect texture and single
phase consistency, are best fulfilled in well aligned
$SmCo_5$- and barium ferrite magnets. There are three unknown parameters of
the theory $n$, $\sigmaf$ and $\sigmas$ which have to be determined by
adjusting the theory to the experiment.
\begin{figure}[htb]
\psfig{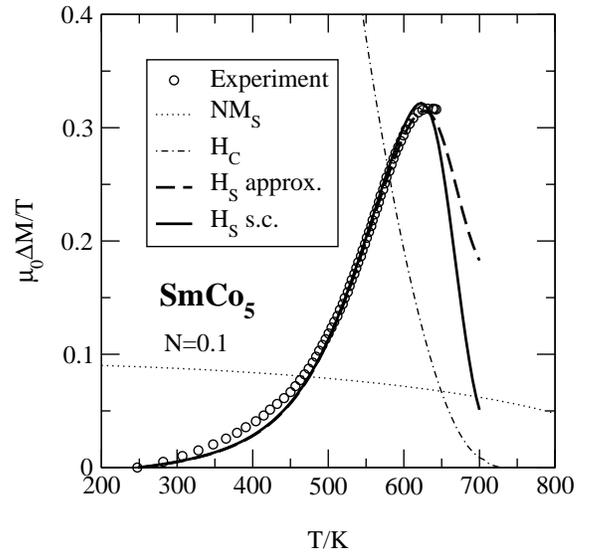}
\caption{Comparison of the best-fits for the $SmCo_5$ sample from Fig. \ref{SmCo5Feld} with $H_S(T)$
calculated self-consistently (solid line) and approximately (broken line) from $H_C(T)$ and
$M_S(T)$.}
\label{SmCo5Hsinfluence}
\end{figure}
Fig. \ref{SmCo5Hsinfluence} shows the best fit results for the $SmCo_5$
sample with N=0.1 used in Fig. \ref{SmCo5Feld} where the (smoothed) measured temperature dependence
of $H_C$ and $M_S$ (shown in Fig. \ref{SmCo5TR}) was used as input.
The best agreement is achieved, if $H_S(T)$ is calculated
self-consistently from eq. (\ref{demagw}). Although up to the maximum the difference
is negligible the approximated formula fails at higher temperatures
where most of the grains are weak.
Furthermore the influence of the form of the m-dependence
of $\sigmafhat$ was checked. It comes out that the initial slope is a little bit increased and the maximum
becomes a little bit lower and broader if the function $\sigmafhat(m)$ is chosen wider.
Unfortunately these changes are to less in comparison with the inaccuracies of the present experimental data
that a definite decision can not be made.\\

\begin{figure}[htb]
\psfig{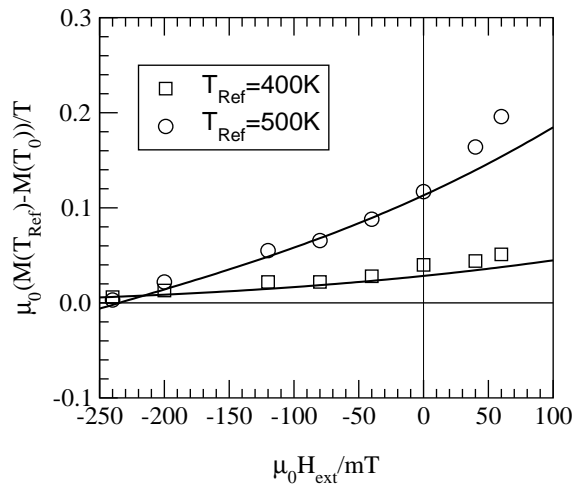}
\caption{Calculated TR between $T_0$ and two reference temperatures $T_{ref}$ in dependence
on a small steady field applied while heating (solid lines). The circles and squares are experimental
points taken from Fig. \ref{SmCo5Feld}.}
\label{SmCo5Nulldurchgang}
\end{figure}
\begin{figure}[htb]
\psfig{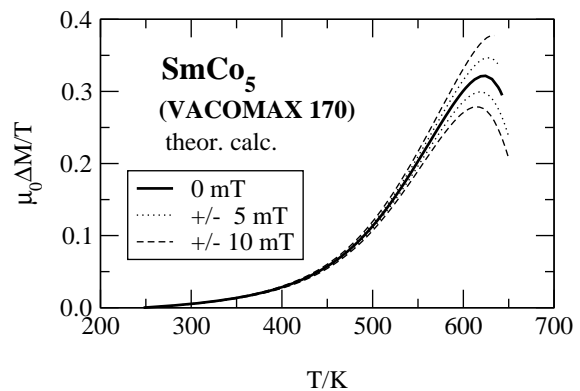}
\caption{ Calculation of the influence of a small external field variation. Solid line: Best-fit TR curve
for the same sample as in Fig. \ref{SmCo5Feld}. Dotted Line: Calculation
for $\mu_0 H_{ext} = \pm 5 \mbox{mT}$. Broken Line: Calculation for a
for $\mu_0 H_{ext} = \pm 10 \mbox{mT}$.}
\label{smco5plusminustheory}
\end{figure}
\begin{figure}[htb]
\psfig{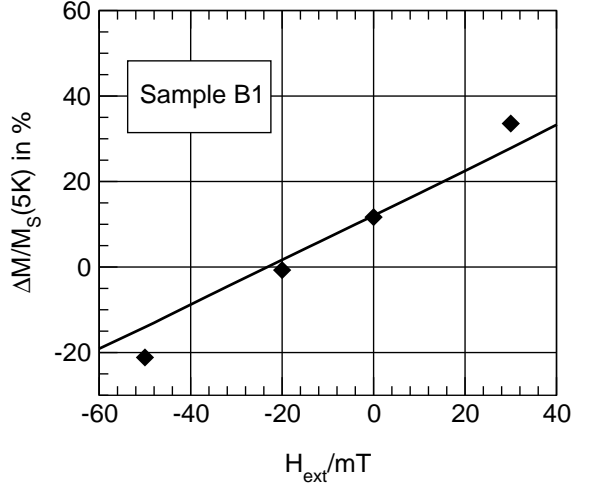}
\caption{Comparison of the calculated (solid line) and measured (diamonds) dependence of the
ITR in barium ferrite on a small steady field applied while cooling from 300 K down to 5 K.}
\label{BaFeNulldurchgang}
\end{figure}

Figs. \ref{SmCo5Feld}
and \ref{SmCo5Nulldurchgang} relate the theory to the experiment. For that the parameters were determined
by fitting the standard experiment, i.e.
with $\hex{2}=0$. The graphs for the other values of $\hex{2}\neq0$ were then calculated
(not fitted!) with the same parameter set. The agreement is fairly well. Of course, it
can be made perfect by fitting the parameter for every curve separately. The variation
of the parameters, given in Tab. \ref{SmCo5Feldfits}, shows that the minima in the least-square sum
are rather flat and that the fit data have to be taken as rough estimates.
\begin{table}[htb]
\begin{center}
\begin{tabular}{|r|c|c|c|}\hline
$H_{ext}$/mT & $\mu_0\sigmas$/T& $\mu_0\sigmaf$/T & $n $ \\ \hline
\multicolumn{4}{|c|}{VACOMAX 170 N=0.1 }\\\hline
+40     & 2.526 &   0.360   & 0.299 \\
  0     & 1.503 &   0.467   & 0.349 \\
-40     & 2.324 &   0.324   & 0.256 \\
-80     & 1.709 &   0.417   & 0.298 \\
-120    & 2.200 &   0.407   & 0.268 \\
-240    & 1.639 &   0.421   & 0.427 \\ \hline
\multicolumn{4}{|c|}{VACOMAX 200 N=0.53 }\\ \hline
  0     & 1.60  &   0.91    & 0.56  \\ \hline
\end{tabular}
\end{center}
\caption{The best-fit values for a VACOMAX 170 sample, determined from the experimental curves
given in Fig. \ref{SmCo5Feld} and for VACOMAX 200 sample \cite{Jahn02a}.
}
\label{SmCo5Feldfits}
\end{table}
From comparing
the curves with other samples of the same material and by calculating the $N$- and $T_0$-dependence with
the values given above we found the fit parameters belonging to $H_{ext}=0$
to be the most reliable ones.
In Fig. \ref{smco5plusminustheory} the theoretical curves
are calculated in analogy to the measurements presented in
Fig. \ref{smco5plusminusexp}.
From the comparison of both figures the qualitative agreement is ocularly,
but the calculated susceptibility is higher than the measured, what may be caused
by the model assumption of ideal mobility of the domain walls within the weak
grains. A further reason is the existence of a small volume fraction of lower coercivity components,
which results in a small dip of the demagnetization curve, i.e.
in a second broad flat peak of the $H_s$ distribution, which is not altered with temperature \cite{Jahn90}.
That weak grains are responsible for the susceptibility
enhancement becomes clear if one calculates the number of weak grains in
dependence on the temperature (dotted line in Fig. \ref{smco5susze}).
For the ITR in barium ferrite multi-domain grains are of less importance, since
upon cooling the coercivity is not reduced below $nM_S$ if n is small. Thus most of the grains
remain in SDS. Again we find an excellent agreement of the
fitted curve with the experimental values, as may be seen from Fig. \ref{BaFerriteN}.
From Tab. \ref{BaFerriteNfits}
\begin{table}[hbt]
\begin{center}
\begin{tabular}[t]{|c|c|c|c|c|}\hline
sample  & $N$  & $\sigmas$/T & $\sigmas$/T & $n$   \\\hline \hline
  B1    & 0.1  &  .124       &  .128       &  .156 \\\hline
  B2    & 0.1  &  .124       &  .129       &  .156 \\\hline
  B3    & 0.3  &  .120       &  .106       &  .153 \\\hline
  B4    & 0.7  &  .131       &  .086       &  .067 \\\hline
  B5    & 0.9  &  .102       &  .113       &  .049 \\\hline
\end{tabular}
\end{center}
\caption{\label{cap:Nfits} Best fit values for the barium ferrite samples from Fig.
         \ref{BaFerriteTR} (B1) and Fig. \ref{BaFerriteN} (B2-B5).}
\label{BaFerriteNfits}
\end{table}
The agreement is much better for the elongated samples, whereas for the prolate discs
internal deviations from
the assumed uniform magnetization seem to be more important \cite{Pastushenkov95}.
In Fig. \ref{BaFeNulldurchgang} the calculated field dependence of the maximum
TR for sample B1 is shown, where the fit values from Tab. \ref{BaFerriteNfits} have been used.

\begin{figure}[htb]
\psfig{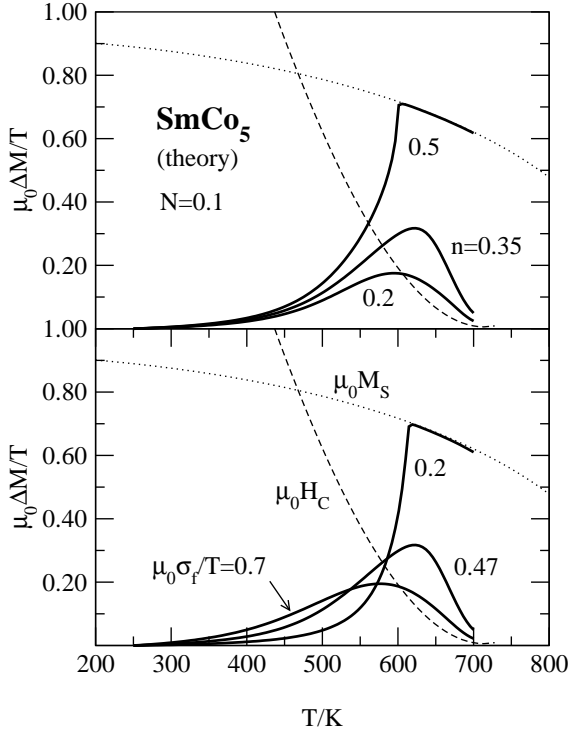}
\caption{Calculation of the influence of $\sigmaf$ and $n$ on the TR. The other parameters are
the best-fit values from Tab. \ref{SmCo5Feldfits}}
\label{SmCo5ParameterInfluence}
\end{figure}

\begin{figure}[htb]
\psfig{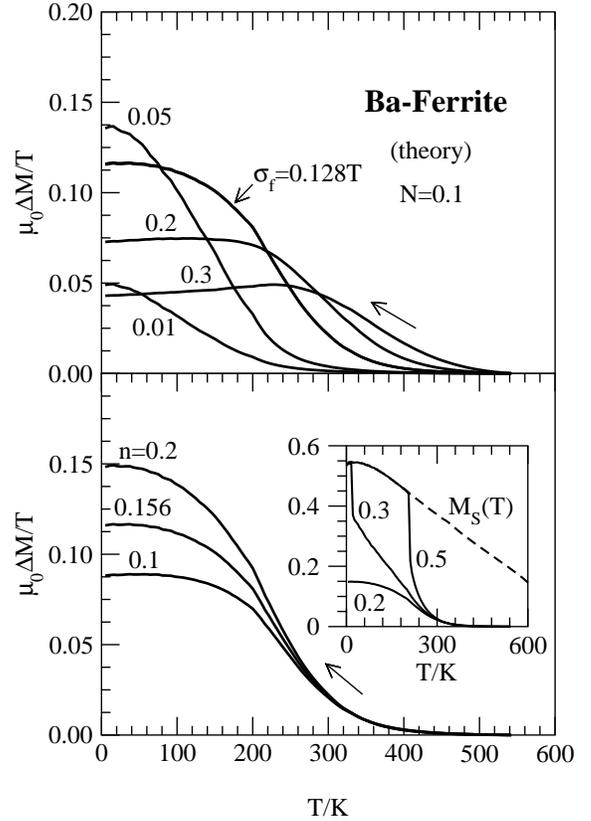}
\caption{Calculation of the influence of $n$ and $\sigmaf$ on the ITR. For the other parameters
the best-fit values for sample B1 from Tab. \ref{BaFerriteNfits} were used.}
\label{BaFerriteParameterInfluence}
\end{figure}

The influence of the intrinsic model
parameters $\sigmas$, $\sigmaf$ and $n$ on the TR is studied
in Fig. \ref{SmCo5ParameterInfluence} for model parameters taken from Tab. \ref{SmCo5Feldfits}
The most interesting features are the strong dependence on the
field fluctuation width $\sigmaf$ and on the internal demagnetization
factor $n$. For small internal demagnetization factors, i.e. for elongated grains, the TR
is small, whereas effects up to 100 \% are possible for platelet-shaped grains. This
corresponds well with the simulation results of Ref. \cite{Zaytzev89}.
The shift of the TR peak to higher temperatures as well as the sharpening with decreasing
fluctuation width can be understood easily from the fact that the switching field of a volume
fraction with given $H_s$ has to be reduced more by temperature before it contributes to the
TR. Taking into account the different temperature dependence of $H_C$ the same may be said
for the ITR in barium ferrite, as seen in Fig. \ref{BaFerriteParameterInfluence}.
\section{Discussion}
Summarizing the experimental facts we find that TR or ITR
is observed in all technical relevant
permanent magnetic materials. Regarding the prerequisites for TR and ITR we found
that a necessary condition for the effect is the preparation
of an asymmetrical demagnetized initial state, where the average switching field of positively and
negatively magnetized grains is different. A further condition is the existence of
a sufficient broad distribution of field fluctuations. Furthermore the reduction
of the coercivity with temperature has to be large enough, that the interaction fields may overcome
the switching fields. The sign of the temperature coefficient of coercivity determines
whether normal TR or ITR happens.
Some features are common to all the four considered groups of permanent magnets (i) the strong
dependence on the demagnetization factor $N$ of the sample (ii) the extreme sensibility against
a small superimposed steady-field $H'$ (iii) the strong dependence on the initial temperature $T_0$
(iv) the increase of the susceptibility after a TR cycle and
(v) the reproducibility of TR in "repeating experiments".
Otherwise, looking in detail, every permanent magnetic material shows some individual characteristics.
For the $SmCo_5$ magnets it is mainly the survival of a TR effect beyond the temperature $T_{Hc}$, due
to the big difference $T_C-T_{Hc}$. For the $NdFeB$ magnets it is the very small
switching field distribution and the decoupling of the magnetic
active grains which reduce the effect. The peculiarities of the barium ferrite magnets are caused by the
different temperature dependence of the coercivity. The $Sm_2Co_{17}$-type magnets differ
from all the others by the underlying coercivity mechanism of the main magnetic phase.
Regarding the latter materials, it is undoubtedly that the samples, where measurements were published,
exhibit TR with all the main features. Otherwise the magnitude of the effect is so small, that it
is at present impossible to decide, whether pinning controlled magnets really show TR.
Moreover, also within a given material group, samples produced with different technologies may
differ considerably in the amount of TR.\\

Nevertheless most of the observed features can be well understood in the framework of
the above presented theory, at least qualitatively, if it is possible to estimate how
the special characteristics of a material are reflected in the model parameters. The latter
are mainly the internal demagnetization factor $n$ and the width of the field fluctuation $\sigmaf$.
The parameter $\sigmas$ results from the assumption of a gaussian SFD, which was introduced for
simplicity. This simplification is of less importance if one calculates the TR, since it is very
sensitive to field fluctuations but less sensitive to the SFD. For the hysteresis curve the situation
is vice versa.
Thus a single gaussian may not be accurate enough to fit the hysteresis curve in good
quality. In principle this situation may be improved by calculating a more realistic SFD
from the  measured hysteresis curve. Regarding the calculation of the TR this
would result in substituting the erf-function by the adequate primitive of the SFD.
We do
not think that this is the main point, instead we regard the simple version of
the "inclusion approximation" ( for details see \cite{Schumann01}) as the most
problematic approximation, since it describes the retroaction of the
different environments onto  the different grains by only one parameter, the
internal demagnetization factor $n$. An argument in favour of this approximation is
the fact, that it yields exactly the same relation between the coercivity
and the related (averaged) switching field as introduced both empirically \cite{Hirosawa86}
and theoretically \cite{Kronmueller88}. Through this paper we used the term
``grain'' for the magnetic bistable
units for simplicity. Indeed, these magnetic units may be different from the
crystallites. Especially if there are strong interactions between the grains, it
may be that some grains agglomerate and switch coherently. For instance the small internal
demagnetization
factor $n$ in the above cited barium ferrite samples shows that the magnetic units are elongated, so it
seems possible, that short chains of crystallites play the role of the magnetic
units.
The other approximations are of less
importance. Of course, the magnetic field fluctuates in three dimensions
regardless of the assumed ideal texture, but, in the case of strong anisotropy it
is merely the z-component of the field which switches the grains, and these
fluctuations will be Gaussian again and are therefore contained in the model.
The latter is supported by the TR and ITR experiments at isotropic sintered
samples of $SmCo_5$ and barium ferrite. Both the remanence and the z-component of the
fluctuating fields are reduced by a factor two with respect to the aligned samples,
This yields approximately the observed factor four for the reduction of the
TR and ITR resp..
The theory works best for the well textured $SmCo_5$- and barium ferrite samples.
For $NdFeB$, albeit curves may be fitted as well, the meaning of the fit parameters
is not as clear. The existence of nonmagnetic phases will reduce the effective
magnetization, i.e. the magnetic units may be thought as a composition of a magnetic
nucleus surrounded by a small nonmagnetic layer \cite{Kronmueller91}. Thus the mean
field can vary between $\pm c M_S$, with $c$ being the content of magnetic phases, whereas
in the switching conditions, it must be taken into account, that the switching unit is the magnetic
nucleus only. Furthermore, the normalization of the SFD has to be changed to $c$.
Without any calculation it may be understood, that the TR in $NdFeB$ is reduced in comparison
with $SmCo_5$. One reason is the smaller switching field distribution, which reduces the
asymmetry in the dc-demagnetized initial state. Furthermore, the nonmagnetic layer
increases the distance between the magnetic
grains and therefore reduces the divergence of the magnetization, what results in smaller
field fluctuations.
If one does not ascribe the observed TR in $Sm_2Co_{17}$ to imperfections, one has to
take into account the cell structure yielding the volume pinning mechanism.
If a grain of such a magnet is switched, the domain
wall moves by successively switching these cells, i.e. as switching units one has to regard
now the cells, not the grains. Thus, also "single crystals" of $Sm_2Co_{17}$ should exhibit TR.
This was indeed observed \cite{Ivanov91}. If the precipitation structure is perfect, all
the cells will exhibit the same switching field, resulting in a very small switching field
distribution, what reduces the TR considerably. The switching of adjacent cells
will be very correlated, resulting in a movement of the domain wall.
Furthermore the exchange coupling of neighboring cells
will surely dominate over the dipole-dipole interaction due to the smallness of the cells.
Although this may be partly taken into account by allowing for internal demagnetization
factors $n$ bigger than one, the averaging over the neighboring grains as was done by
the inclusion approximation becomes very questionable. Thus, the presented theory may be not very
suited for the case of $Sm_2Co_{17}$.

Regarding the data fitting we learned from Fig. \ref{approximations} that both
weak grains and down-switching of hard grains for higher temperatures has to be taken into account.
This increased the computational effort considerably, due
to the self-consistency requirements for $\bar{H}_s(T)$ and $\mean{M}$.
The model parameters determined from the fitting procedure have to be considered as
estimates. To increase their reliability improvements of both the theory, as outlined above,
and the experimental data are necessary.
Regarding the experiment, both labor experience and the presented theory show that the most critical
point in a TR experiment is the sample demagnetization factor. Of course, a closed circuit measurement would be highly
desirable, as was shown in the very first reported TR experiment. This is
hard to realize, especially for high-coercive samples and at low temperatures. Therefore rotational
ellipsoidal samples with the symmetry axis along the texture axis should be used to secure
the homogeneity of the mean internal field. A second point is a very accurate measurement of
$H_C(T)$ and $M_S(T)$. Especially the coercivity should be measured before and after a TR
experiment, where measurements beyond $T_{max}$  should be the last measurements at all, since
they may result in irreversible changes within the material. Furthermore the field dependence of the
irreversible susceptibility should be determined for several temperatures, since this
would allow to determine an estimate of the SFD and to proof the assumption that the temperature dependence
of the switching fields of different grains are equal. Since the theory was extended to arbitrary
starting points, small deviations from the dc-demagnetized state are no problem. On the contrary,
different small steady fields should be applied during the TR experiment to determine the suppression field
$H'$, as an experimental estimate of the field fluctuation width.\\

For the theory of the TR and ITR it was inevitable to develop a
theory for the hysteresis and the recoil curves also.
Whereas there exist a multitude of micro-magnetic
calculations trying to explain the coercivity by assuming special
micro-structures which give rise to either pinning of domain walls or
nucleation, our theory starts a little above this level, since we do not
judge about the reason for a switching field, but simply accept its existence.
In this point the theory resembles the Preisach \cite{Preisach35} model (PM).
A theory of the TR based on the classical model is not able
to explain TR effects of more than 50\%. Furthermore this model was shown to have the congruency and
wiping-out property \cite{Mayergoyz86}, which is a severe disadvantage, since
in most of the interesting cases the measured magnetization behavior violates
these conditions.
The main disadvantage of the classical Preisach model is due to the fact that the
Preisach distribution function is not dependent on the magnetization state
\cite{Preisach35,DellaTorre99}.
To describe TR effects up to 100\% it was necessary to
take into account many-particle effects, i.e. the feedback of the neighborhood
onto the grain, which result in
a dependence of the form of the fluctuation field distribution on the magnetization
state \cite{Schumann87}. As shown above the TR model was further completed step by
step by taking into account the contribution of multi-domain particles and by the
inclusion approximation. Parallel in time and mainly in the context of recording media
the classical PM was improved also by taking into account a mean field
shift \cite{DellaTorre66,Basso94} in the moving PM (MPM).
Although at first glance the ''moving term'' $H+k\mean{M}$
looks very similar, it describes a rigid mean field shift which is the same for all magnetic units.
Contrary, the "inclusion correction" $H-n(\sigma M_S-\mean{M})$)describes the deviation of
the grain magnetization from the magnetization of its neighborhood.
To take reversible processes into account the MPM was modified another time
by allowing non-rectangular elementary hysteresis loops \cite{DellaTorre94} of the magnetic units,
or by splitting the Preisach density function, as it was done in
\cite{Song00,Stancu01} (GPM). The common drawback of all the mentioned Preisach
models is the neglect
of the state and history dependence of the {\em form} of the interacting field
distribution. This had been shown already by one of the first computer
simulations \cite{Moskowitz67}, and also by experiments
\cite{PardaviHorvath94}. The variable variance Preisach model accounts partly
for that observation \cite{Vajda93}. If the improvements of the various mentioned
Preisach models would be mounted together, it may be probably
possible to describe the TR in permanent magnets also on that basis,
if reasonable assumptions regarding the
temperature dependence of the model parameters are added.
A first step in this direction was done in Ref. \cite{Mueller2000}.

In conclusion it can be said that the investigation of the TR and ITR may be used to
characterize modern permanent magnet materials. In connection with the presented model it provides
an alternative tool for studying interaction effects in relation to the temperature
dependence of the saturation magnetization and switching field distribution.

\end{document}